\documentclass[acmsmall,nonacm]{acmart}  %% arXiv single-file variant -- GENERATED by make-arxiv.sh, do not hand-edit

\setcopyright{none}
\acmJournal{TAAS}
\acmYear{2026}
\acmVolume{0}
\acmNumber{0}
\acmArticle{0}
\acmDOI{}

\usepackage{mathtools}
\usepackage{booktabs}
\usepackage{multirow}
\usepackage{array,threeparttable}
\usepackage{tikz}
\usepackage{pgfplots}
\pgfplotsset{compat=1.18}
\usepackage{algorithm}
\usepackage{algpseudocode}
\usepackage{pifont}
\usepackage{bm}
\usepackage{subcaption}
\usepackage[capitalize]{cleveref}

\AtEndPreamble{%
  \theoremstyle{acmdefinition}%
}

\title[Autonomic Federated-Market Orchestration for the Edge--Cloud Continuum]{%
  Autonomic Federated-Market Orchestration for the Edge--Cloud Continuum}

\author{Lauri Lov\'{e}n}
\orcid{0000-0001-9475-4839}
\authornote{Corresponding author.}
\affiliation{%
  \institution{Future Computing Group, University of Oulu}
  \city{Oulu}
  \country{Finland}
}
\email{lauri.loven@oulu.fi}

\author{Roberto Morabito}
\affiliation{%
  \institution{EURECOM}
  \city{Sofia Antipolis}
  \country{France}
}

\author{Abhishek Kumar}
\affiliation{%
  \institution{University of Jyväskylä}
  \city{Jyväskylä}
  \country{Finland}
}

\author{Susanna Pirttikangas}
\affiliation{%
  \institution{AMD Silo AI}
  \city{Oulu}
  \country{Finland}
}

\author{Jukka Riekki}
\affiliation{%
  \institution{Future Computing Group, University of Oulu}
  \city{Oulu}
  \country{Finland}
}

\author{Sasu Tarkoma}
\affiliation{%
  \institution{University of Oulu and University of Helsinki}
  \city{Oulu / Helsinki}
  \country{Finland}
}

\renewcommand{\shortauthors}{Lov\'{e}n et al.}

\ccsdesc[500]{Computer systems organization~Self-organizing autonomic computing}
\ccsdesc[500]{Computer systems organization~Distributed architectures}
\ccsdesc[300]{Computing methodologies~Multi-agent systems}
\ccsdesc[300]{Theory of computation~Algorithmic mechanism design}

\keywords{Autonomic computing, edge--cloud continuum, self-organisation, market-based allocation, federated brokers, MAPE-K, sovereignty enforcement, Walrasian equilibrium, polymatroidal feasibility}

\begin{document}

\begin{abstract}
The edge--cloud computing continuum demands self-management across autonomous administrative domains while honouring tenant- and operator-specified data sovereignty. We present \emph{Neural Pub/Sub}, a federated-broker autonomic substrate whose self-organising behaviour emerges from market-based price signals rather than centralised control. Each broker closes a MAPE-K loop over health and load monitoring, marginal-cost clearing-price analysis, placement over a polymatroidal feasibility region, federated cross-domain dispatch, and bounded-staleness peer price signals. The Plan step is anchored in a Walrasian convergence proposition: under gross-substitutes valuations on tree and series-parallel service-dependency DAGs, decentralised price-based allocation matches the welfare of a centralised oracle.

On a 4-VM, 48-worker federated testbed (1005-run campaign with a fair-process-count sharded-oracle comparator), the market beats a single-process oracle by 2--4\,\% (45 of 45 per-seed wins) and stays within $\pm1.5\,\%$ of a four-shard oracle across all nine (pipeline, load) cells. Under saturation, round-robin completion collapses ($98.8\%\to22.4\%\to3.3\%$) while the market preserves completion; the advantage decomposes into three Walrasian properties: information completeness, admission control, and price discovery. Federation withstands broker death and network partition (completion $\geq98.7\%$), and sovereignty enforcement adds no measurable runtime overhead.
\end{abstract}

\maketitle

\section{Introduction}\label{sec:intro}

The edge--cloud computing continuum has become the default substrate for scientific and operational workloads, spanning instruments, edge sites, regional clouds, and centralised HPC facilities under independent administrative control~\cite{parashar2025everywhere}. Its complexity, heterogeneity, and dynamic nature place self-management on the critical path: workloads must be monitored, analysed, planned for, and dispatched without a human in every loop, and without a single coordinator with global visibility into all domains. This is the autonomic-computing problem~\cite{kephart2003vision,ibm2003autonomic} restated for the continuum: a federation of autonomous administrative domains, each governed by its own operator, must self-organise to deliver multi-stage AI pipelines whose stages are placed across the continuum subject to capacity, latency, and sovereignty constraints.

\begin{figure}[t]
    \centering
    \includegraphics[width=0.7\linewidth]{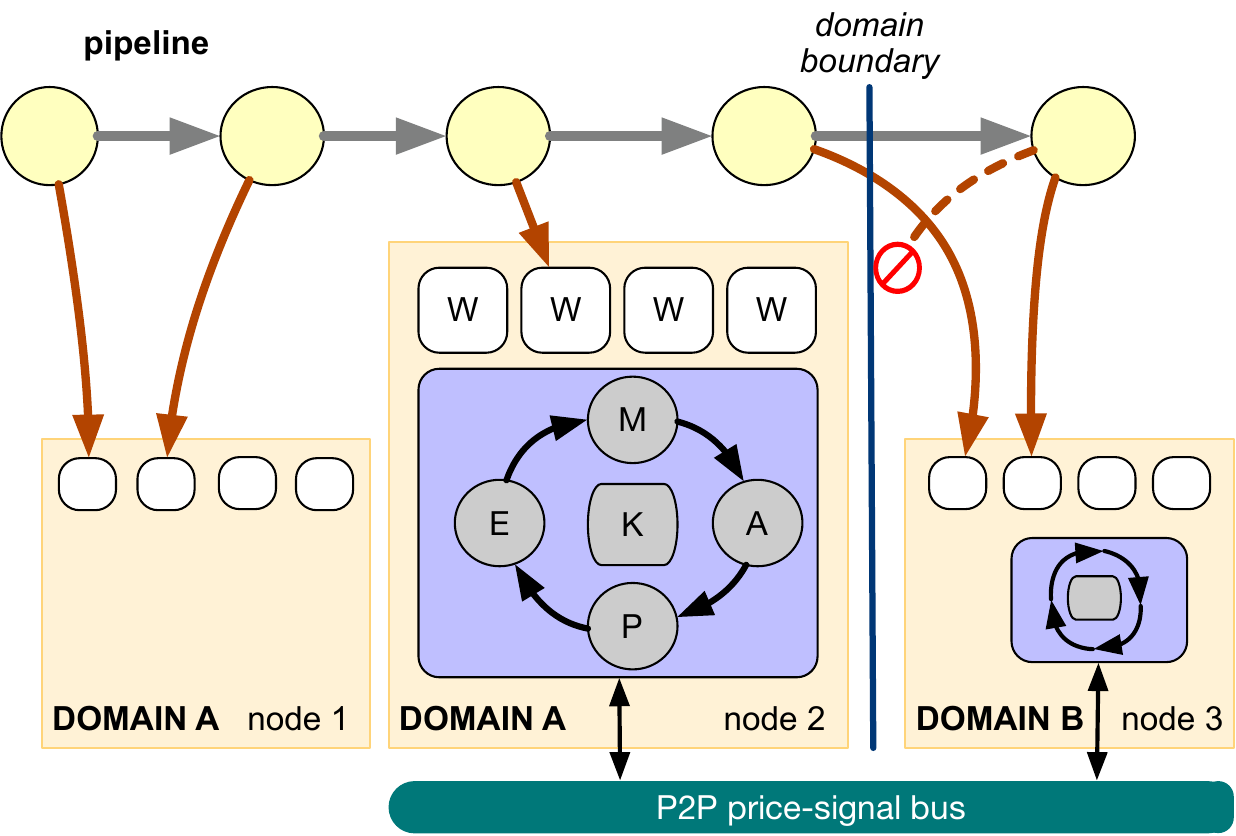}
    \caption{Conceptual overview of Neural Pub/Sub. A pipeline (gray) is
    dispatched stage-by-stage (orange) onto workers across
    administrative \emph{domains}; each domain runs one \emph{broker} closing a
    MAPE-K loop (expanded for Domain~A). Brokers coordinate peer-to-peer over
    clearing-price signals (teal bus), without a central coordinator;
    \texttt{local-only} stages cannot cross a \emph{domain boundary} ($\oslash$).
    Schematic of the mechanism (\cref{sec:DistributionArch}); not the evaluation
    testbed (\cref{sec:Evaluation}).}
    \label{fig:overview}
\end{figure}

We address this problem with a federated-broker substrate, \emph{Neural Pub/Sub}~(\cref{fig:overview}), in which self-organisation is realised through market-based price signals rather than centralised coordination.
 Each domain runs a broker that monitors local workers, publishes a clearing-price summary to its peers, and dispatches stages of incoming pipelines either locally or across a federation overlay according to a price-based trade rule. The substrate's autonomic behaviour closes a MAPE-K control loop~\cite{kephart2003vision,ibm2003autonomic} (\cref{sec:mapek}): per-broker health checks (Monitor), marginal-cost clearing-price computation (Analyse), placement decisions over a polymatroidal feasibility region (Plan), federated cross-domain dispatch (Execute), and shared peer subscription summaries with cached price signals as Knowledge. The continuum is instantiated as four edge-to-cloud administrative domains; in our testbed these correspond to O-RAN logical components, but the substrate is O-RAN-agnostic and applies to any four-domain edge-cloud topology.

Existing self-management substrates address fragments of this problem but not its full scope. Container orchestrators (Kubernetes, KubeEdge~\cite{xiong2018kubeedge}) and batch schedulers (Kueue~\cite{kueue2024}, Volcano~\cite{volcano2024}) operate within a single administrative domain with a single control plane that has full worker visibility, foreclosing federation across vendor or jurisdictional boundaries. Stream-processing systems (Apache Kafka, Flink~\cite{carbone2017flink}, Ray~\cite{moritz2018ray}) move data at scale but route by topic partitions, not by semantic content of service requests. ML pipeline orchestrators (Kubeflow~\cite{bisong2019kubeflow}, Argo Workflows, MLflow~\cite{zaharia2018accelerating}) compose multi-stage pipelines but determine pipeline structure at design time and operate within a single cluster, with no mechanism for cross-domain placement, slice-aware QoS, or sovereignty enforcement. Self-adaptive software frameworks~\cite{salehie2009self,weyns2020software,cheng2009software} have established the principles of MAPE-K and feedback-driven adaptation; this paper instantiates those principles for the continuum, with a market-based Plan step grounded in mechanism-design theory.

\paragraph{Theoretical anchor.} The architecture is grounded in a Walrasian convergence proposition imported from companion work~\cite{trilogy_p1}: the AI Service Markets framework~\cite{trilogy_p1} establishes that when service-dependency DAGs have tree or series-parallel structure, the feasible allocation region is polymatroidal and a Walrasian equilibrium exists under gross-substitutes valuations~\cite{kelso1982job,gul1999grosssubstitutes}. Decentralised price-based allocation can therefore, in principle, match the efficiency of a centralised oracle; in practice, at the scales we test, it can match a fair-process-count comparator (\cref{sec:f1}). The proposition motivates the Plan step's mechanism design and is restated self-containedly in Section~S2 of the electronic supplement.

\paragraph{Sovereignty as a first-class self-management constraint.} The substrate enforces tenant- and operator-specified data sovereignty (which stage types may leave their originating domain) as a coordinate-wise constraint on the feasibility region. The motivation is regulatory (GDPR transfers, EU AI Act conformity tiers, NIS2 supply-chain obligations), but the architectural treatment is generic: sovereignty is one instance of the broader self-management requirement that autonomic decisions honour domain-level invariants without external orchestration.

\paragraph{Scope.} The campaign reported here exercises a federated edge-cloud topology of four administrative domains on a single-data-centre telco-edge 4-VM testbed at the 5G Test Network Finland, Oulu (5GTNF)~\cite{piri2016gtn} with a single \texttt{tc qdisc netem}-emulated WAN link. The continuum claim is rather architectural and orchestration-layer than infrastructural: the broker substrate composes self-management across four-domain edge-to-cloud topologies, and the experimental evidence covers the orchestration loop's behaviour under load, failure, partition, and sovereignty constraints. Cross-site WAN deployment, HPC-tier integration, asymmetric-trust / asymmetric-capacity / asymmetric-data-quality regimes, and live agent-runtime integration are out of scope and are discussed as future work in \cref{sec:Discussion}.

\paragraph{Contributions.} This paper contributes the federated distribution architecture for Neural Pub/Sub, the explicit MAPE-K mapping that frames it as an autonomic substrate, and a 1005-run experimental campaign across the edge-cloud continuum supplemented by a fair-process-count sharded-oracle comparator:

\begin{enumerate}
    \item \textbf{Autonomic federated-market substrate (\cref{sec:DistributionArch}).} A four-domain federation of neural brokers that close a MAPE-K loop (\cref{sec:mapek}) over marginal-cost clearing prices, polymatroidal placement, cross-domain dispatch, and peer subscription summaries. The Plan step operationalises Walrasian convergence (Proposition~\ref{prop:walrasian}, citing~\cite{trilogy_p1,kelso1982job,gul1999grosssubstitutes}) for tree and series-parallel DAGs through five engineering refinements that are jointly sufficient for the empirical envelope tested (\cref{sec:market-impl}); per-component leave-one-out attribution is deferred. The substrate is composed as a static-DAG orchestration layer that can sit beneath dynamic agent runtimes (LangGraph~\cite{langgraph2024}, AutoGen~\cite{wu2023autogen}, Anthropic MCP~\cite{anthropic_mcp2024}, Google A2A~\cite{google_a2a2025}); live agent-runtime integration and AI-augmented loop components are deferred.

    \item \textbf{Autonomic decentralisation matches centralised at equal process count and dominates the single-process oracle by 2--4\,\% (\cref{sec:f1}).} The four-broker federated market is 2--4\% lower mean latency than a single-process oracle across every (pipeline, load), with 45 of 45 per-seed wins (sign-test $p\approx 2.8\times 10^{-14}$, Hodges--Lehmann median $-39.6$\,ms, 95\% bootstrap CI $[-44.1,-35.3]$\,ms). A fair-process-count sharded-oracle comparator (4 coordinator processes, designated coordinator pulls peer state and runs find\_placement over a merged topology) keeps the gap within ${\pm}1.5\%$ across all 9 cells: the market dominates linear-chain cqi-chain (3/3, 1.4\,\%), the sharded oracle is marginally favoured at +5\,ms / +2\,ms on anomaly-sp at low / medium load, and ran-entangled ties at the cell level (1 market / 1 sharded / 1 tie). The inversion against the single-process oracle and the within-noise envelope at equal process count together support a sharper claim than ``centralised is impractical'': at equal process count the autonomic substrate trades within $\pm 5$\,ms of an idealised centralised orchestrator while preserving vendor confidentiality, scaling without a single coordinator, and admitting formal Walrasian guarantees.

    \item \textbf{Self-management differentiates only under stress (\cref{sec:f2}).} Under uniform load the market and three heuristic baselines converge to within ${\sim}5$\,ms across every (pipeline, load) cell. Under saturation the conventional round-robin orchestrator's completion rate (CR) collapses $98.8\%\to 22.4\%\to 3.3\%$ across $\lambda\in\{5,10,15\}$\,pps while the market preserves CR. The advantage decomposes into three Walrasian properties (information completeness, admission control, price discovery) absent from non-autonomic baselines.

    \item \textbf{Autonomic robustness under broker death and network partition (\cref{sec:f3}).} The federation withstands broker process kill, emulated edge--cloud network partition, and per-stage worker death across both edge sites: CR\,${\geq}\,98.7\%$ across 75 federation and resilience cells.

    \item \textbf{Sovereignty enforcement adds zero observable runtime overhead (\cref{sec:f4}).} Across 60 governance-grid runs (four enforcement scenarios $\times$ three pipeline types $\times$ five seeds) the mean latency under sovereignty constraints (stages tagged \texttt{local-only} cannot leave the originating domain) matches the unconstrained baseline to within 1--3\,ms, demonstrating that the autonomic substrate honours domain-level invariants without external orchestration cost.
\end{enumerate}

The remainder of this article is organised as follows. \Cref{sec:RelatedWork} surveys autonomic-computing foundations, distributed AI orchestration, semantic communication, pub/sub systems, and AI service marketplaces. \Cref{sec:SystemModel} presents the system model including service-dependency DAGs and the Walrasian convergence result. \Cref{sec:DistributionArch} details the distribution architecture, including the MAPE-K mapping (\cref{sec:mapek}) and sovereignty enforcement (\cref{sec:gov-composition}). \Cref{sec:Evaluation} evaluates the architecture across the 1005-run campaign and the fair-process-count sharded-oracle comparator. \Cref{sec:Discussion} discusses autonomic implications and limitations. \Cref{sec:Conclusion} concludes.
          %% §1 Introduction (autonomic / continuum framing)

\section{Related Work}\label{sec:RelatedWork}

This section positions Neural Pub/Sub against five bodies of work: autonomic computing and self-adaptive systems (\cref{sec:rw-autonomic}), distributed AI orchestration platforms (\cref{sec:rw-orchestration}), publish/subscribe systems (\cref{sec:rw-pubsub}), semantic communication (\cref{sec:rw-semcom}), and AI service marketplaces (\cref{sec:rw-markets}). \Cref{sec:rw-position} closes with a positioning summary.

\subsection{Autonomic Computing and Self-Adaptive Systems}\label{sec:rw-autonomic}

The autonomic-computing programme introduced by Kephart and Chess~\cite{kephart2003vision} and elaborated in IBM's architectural blueprint~\cite{ibm2003autonomic} frames complex distributed systems as collections of autonomic managers, each closing a MAPE-K (Monitor, Analyse, Plan, Execute, Knowledge) control loop over a managed resource. Self-adaptive software extends this with explicit attention to engineering practice: Salehie and Tahvildari~\cite{salehie2009self} survey the landscape of self-adaptation under MAPE-K and adjacent reference models, Cheng et al.~\cite{cheng2009software} lay out a research roadmap for software engineering of self-adaptive systems, and Weyns~\cite{weyns2020software} integrates the contemporary toolbox (architecture-based adaptation, statistical model checking, learning-enabled adaptation) into a single textbook treatment. Parashar~\cite{parashar2025everywhere} reframes the autonomic problem for the edge--cloud continuum, identifying decentralised self-management as the load-bearing capability for science workloads spanning instruments, edge sites, and HPC centres.

Neural Pub/Sub instantiates these principles concretely. Its MAPE-K control loop (\cref{sec:mapek}) closes over per-broker health monitoring, marginal-cost clearing-price analysis, polymatroidal placement planning, federated dispatch execution, and peer subscription summaries with bounded-staleness price signals as Knowledge. The Plan step is anchored in mechanism-design theory rather than feedback heuristics, distinguishing this work from the bulk of self-adaptive software that adapts via control-theoretic or rule-based loops without formal welfare guarantees. The continuum-level framing aligns directly with Parashar's autonomic vision~\cite{parashar2025everywhere}: domains are administratively independent, no single coordinator has global visibility, and self-organisation is the only mechanism by which heterogeneous, dynamic, multi-tenant workloads can be placed at scale. Saleh et al.'s MemIndex~\cite{saleh2025memindex} addresses an adjacent autonomic problem in the same LM-based multi-agent pub/sub family: an adaptive, autonomous distributed memory-management layer that lets agents self-negotiate memory operations through an intent-indexed bipartite graph. MemIndex sits at the agent layer (each LM agent's localized, dynamically updated memory of past user intents) while Neural Pub/Sub sits at the broker layer (each broker's bounded-staleness view of peer subscriptions and prices); a fully integrated autonomic substrate would compose the two---LM agents using MemIndex for distributed memory while their pipelines are placed by the federated brokers of \cref{sec:DistributionArch}.

\subsection{Distributed AI Orchestration}\label{sec:rw-orchestration}

Orchestrating AI workloads across heterogeneous infrastructure has produced three families of systems. Telecom-orchestration standards (NFV-MANO~\cite{etsinfvman001}, ZSM~\cite{etsizsm002}, ONAP~\cite{onap2024}, O-RAN SMO~\cite{oranwg1oad,oranwg2nrtric}) describe the layers below (infrastructure lifecycle) and above (intent and exposure) autonomic placement substrates; Neural Pub/Sub is the autonomic substrate, complementary to but distinct from these management envelopes.

\textbf{Container orchestration.}
Kubernetes and its edge extensions (KubeEdge~\cite{xiong2018kubeedge}, K3s) manage container lifecycles, autoscaling, and scheduling. KubeEdge supports up to 100\,000 edge nodes and provides offline operation with cloud synchronisation~\cite{xiong2018kubeedge}. Modern batch schedulers Kueue~\cite{kueue2024} and Volcano~\cite{volcano2024} add priority queues, resource quotas, gang-scheduling, and admission control. All of these treat workloads as opaque containers within a single Kubernetes control plane: scheduling is based on resource requests and affinity labels, not the semantic content of the data flowing through the pipeline, and federation across administrative or vendor boundaries is not addressed.

\textbf{Stream processing and data pipelines.}
Apache Kafka provides durable, partitioned event streams; Apache Flink~\cite{carbone2017flink} adds stateful stream processing with exactly-once semantics; and Ray~\cite{moritz2018ray} offers a unified runtime for data processing, training, and serving with actor-based parallelism. Recent versions of Flink (2.2) introduce native ML inference operators and async LLM calls within streaming pipelines. These systems excel at high-throughput data movement and transformation but route events by topic partitions or explicit key assignments, not semantic content.

\textbf{ML pipeline orchestration and serving.}
Kubeflow Pipelines~\cite{bisong2019kubeflow}, Argo Workflows, and MLflow~\cite{zaharia2018accelerating} compose multi-stage AI pipelines as developer-defined DAGs on Kubernetes; KServe~\cite{kserve2024} and Triton~\cite{triton2024} serve individual models with autoscaling and dynamic batching. These share with Neural Pub/Sub the DAG abstraction for multi-stage workflows but differ in three respects: (i)~pipeline structure is determined at design time by developer logic, not composed dynamically at runtime based on the semantic content of service requests; (ii)~scheduling operates within a single Kubernetes cluster (one administrative domain), with no federation protocol or cross-domain placement; and (iii)~placement decisions are based on resource requests and affinity labels, not on slice-level QoS constraints or governance policies. Neural Pub/Sub addresses a complementary problem: not how to execute a known pipeline within a managed cluster, but how to discover, compose, and place pipelines across autonomous domains based on the semantic content of service requests.

\textbf{LLM-based agentic orchestration.}
LLM-based multi-agent frameworks~\cite{guo2024llmmultiagent} mark a static-to-dynamic transition at the task orchestration layer: statically declared graphs in AutoGen~\cite{wu2023autogen} and CrewAI~\cite{crewai2024} versus runtime-composed graphs in LangGraph~\cite{langgraph2024} and the OpenAI Agents SDK~\cite{openai_agents2025}, with typed cross-framework surfaces such as Anthropic MCP~\cite{anthropic_mcp2024} and Google A2A~\cite{google_a2a2025}. All of them compose \emph{agent} graphs at the task layer; Neural Pub/Sub composes \emph{resource-placement} graphs at the orchestration layer, beneath the agent runtime. Saleh et al.~\cite{saleh2025messagebrokers} survey broker evolution for GenAI agents but identify no formal allocation mechanisms; model routers (OpenRouter~\cite{openrouter2025}, LiteLLM~\cite{litellm2024}, RouteLLM~\cite{routellm2024}, \cite{notdiamond2024}) select endpoints by cost or latency without incentive compatibility or governance. \emph{Scope.} \Cref{sec:DistributionArch}'s mechanism requires the placement-call DAG to be known at allocation time, the regime of current orchestration substrates (Kubeflow, Argo, Airflow). Runtime DAG growth (e.g., LangGraph or A2A multi-step delegation) is supported by issuing one placement call per growth step; per-call welfare guarantees carry over but not as a single global guarantee across step sequences, since A2A-style multi-step delegation is non-monotonic-composition (a later step may revoke or rewrite earlier sub-task assignments) and the per-call Walrasian guarantees do not compose across such revocations. Dynamic-graph allocation is future work.

\textbf{Computing-continuum scientific-workflow orchestration.}
The Edge--HPC/Cloud continuum vision originates in the in-transit and in-network computing programmes of Beckman et al.~\cite{beckman2020harnessing} and is reframed as an autonomic-management challenge by Parashar~\cite{parashar2025everywhere}. Workflow systems for the science-end of the continuum --- Pegasus~\cite{deelman2015pegasus}, Parsl~\cite{babuji2019parsl}, Swift~\cite{wilde2011swift}, CometCloud~\cite{kim2009cometcloud} --- compose multi-stage scientific pipelines across edge instruments, intermediate analysis tiers, and HPC centres. They share Neural Pub/Sub's pipeline-as-DAG abstraction and the cross-tier-placement question, but their decision substrate is centralised (a single workflow planner with global view), and welfare guarantees are not the design objective. In-transit computing positions the data path itself as a placement target~\cite{beckman2020harnessing}; the Neural Pub/Sub substrate is complementary, operating at the orchestration layer above the data plane. \emph{The HPC tier of the SI's title is acknowledged but not exercised in the present campaign}: the testbed is edge-cloud (4 administrative domains, 5GTNF), and HPC-tier integration --- e.g., a Pegasus-style cross-DC workflow with a tier-3 supercomputing site --- is deferred to follow-on work. The autonomic mechanism (federated brokers, market-based clearing, MAPE-K loop) is HPC-tier-compatible by design but unproven in that regime.

\textbf{Gap.}
The orchestration field has undergone a static-to-dynamic transition at the task layer (Airflow $\to$ LangGraph) but not at the resource layer (Kafka and Kubernetes remain static, single-cluster, semantically opaque), and the continuum-orchestration substrates above (Pegasus, Parsl, Swift, CometCloud) are centralised single-planner systems without welfare guarantees or federation across administrative boundaries. Neural Pub/Sub provides the resource-layer transition \emph{with} formal guarantees, instantiating the autonomic-computing principles of~\cite{kephart2003vision,ibm2003autonomic,salehie2009self,cheng2009software,weyns2020software,parashar2025everywhere} in a federated, market-based substrate.

\subsection{Publish/Subscribe Systems}\label{sec:rw-pubsub}

Publish/subscribe (pub/sub) decouples communicating endpoints in space, time, and synchronisation~\cite{eugster2003many}. Topic-based systems (Kafka, MQTT) route by partition or channel name; content-based systems evaluate filters over message attributes~\cite{tarkoma2012publish}. Carzaniga et al.~\cite{carzaniga2001design} introduced SIENA, a wide-area event notification service with predicate-based filters and an overlay routing algorithm that propagates subscriptions to minimise unnecessary forwarding. Subsequent work extended content-based pub/sub to structured P2P overlays~\cite{tarkoma2012publish}, distributed hash tables, and elastic cloud deployments. These systems evaluate filters using Boolean predicates, attribute ranges, or regular expressions over structured fields. When subscriptions and events are expressed in unrestricted natural language (or carry multimodal content), paraphrases, synonyms, and contextual meaning escape predicate-based filters. Neural Pub/Sub replaces predicate evaluation with embedding-based clustering (the matching layer is exercised in this paper as a frozen, pre-calibrated component; see \cref{sec:neural-router-recap}) and propagates subscription \emph{summaries} (centroid embeddings and cluster radii) rather than raw subscription text, enabling cross-domain routing in embedding space.

\subsection{Semantic Communication}\label{sec:rw-semcom}

Semantic communication transmits the \emph{meaning} of messages rather than their bit-exact form, moving the design objective from Shannon source/channel coding to goal-oriented effectiveness~\cite{strinati2021semantic,qin2022semantic}. Deep semantic codecs (DeepSC~\cite{xie2021deep}, DeepJSCC~\cite{bourtsoulatze2019deep}) use Transformer-based joint source-channel coding under low-SNR conditions, and recent LLM-aware extensions~\cite{wang2024semantic,letaief2022edgeai6g} couple semantic codecs with edge computing. These approaches optimise transport for the semantics of transmitted content; Neural Pub/Sub operates one layer up, on already-decoded content, deciding \emph{which receivers} should act on it. The two layers compose: \cref{sec:semcomm-positioning} sharpens the boundary by framing embedding-based matching as a routing-oriented rate-distortion problem.

\subsection{AI Service Marketplaces and Resource Allocation}\label{sec:rw-markets}

The continuum's multi-stakeholder nature motivates market-based approaches to resource allocation.

\textbf{Service-function-chain placement and virtual-network embedding.} Pipeline placement is closely related to \emph{service-function-chain (SFC)} placement~\cite{bari2016sfc} and \emph{virtual-network embedding (VNE)}, both NP-hard with ILP and combinatorial-auction reductions. The combinatorial-auction angle inherits the Lehmann--Lehmann--Nisan / Nisan--Segal hardness picture~\cite{lehmann2006combauct,nisan2006communication}: without gross-substitutes structure, welfare-maximising allocation in combinatorial auctions is communication-complexity-hard and computationally NP-hard, justifying the integrator-encapsulation route to a GS-compatible representation rather than direct ILP. Neural Pub/Sub differs from classical SFC/VNE in three ways: (i)~valuations are gross-substitutes by construction at the agent-facing layer (after integrator encapsulation), making price decomposition into a Walrasian mechanism well-defined; (ii)~placement is decentralised across federated brokers using only summary information, trading optimality for vendor confidentiality; and (iii)~governance constructs (\cref{sec:gov-composition}) are absent from classical SFC/VNE formulations.

\textbf{Classical DAG-scheduling baselines.} List-scheduling heuristics on static DAGs (HEFT~\cite{topcuoglu2002heft}, PEFT, CPOP) target makespan minimisation under a fixed task graph and a known set of heterogeneous processors with deterministic compute and communication costs~\cite{kwok1999survey,graham1966bounds}. Neural Pub/Sub addresses a different problem class: pipelines arrive as a Poisson stream, the resource pool is shared across continuous arrivals, the allocation criterion is welfare (combining latency, queue depth, sovereignty, and federation cost) rather than makespan of a single DAG, and the decision is decentralised across federated brokers exchanging only price signals. HEFT-style baselines do not directly compose with a continuous-arrival welfare objective: makespan is undefined when the workload is open-ended, and HEFT's centralised global-graph view contradicts the federated-broker constraint. Their absence from the comparison is therefore an honest scope choice, not an oversight; the centralised single-process oracle and fair-process-count sharded oracle play the role of the welfare-maximising upper-bound baselines that classical DAG-scheduling literature would call for.

\textbf{Edge resource markets and placement strategies.}
Auction-based mechanisms for edge resource allocation have been studied under various settings: truthful mechanisms for multi-attribute demand in mobile edge computing~\cite{zhang2024truthful}, pricing strategies for edge-cloud service provision, and RL-based task offloading formulated as Markov decision processes. These approaches allocate individual resources but do not model the dependency structure of multi-stage AI pipelines. At the placement level, common reallocation strategies (proximity-based, load-balanced, and random) exhibit fundamentally different trade-offs between latency minimisation and load distribution~\cite{loven2021weathering}. Under high load, load-sensitive strategies can trigger cascading superfluous reallocations (``reallocation storms'') where up to 15--20\% of all tasks are reallocated needlessly~\cite{loven2022stormy}; simpler random strategies avoid this phenomenon because they do not react to load signals. This trade-off is directly relevant to our market-based mechanism, which uses price signals as a coordination layer between load-sensitive and proximity-aware placement (\cref{sec:market-allocation}). The Walrasian equilibrium framework~\cite{kelso1982job,gul1999grosssubstitutes} provides the theoretical foundation for price-based allocation under gross-substitutes valuations; Ausubel's ascending clinching auction~\cite{ausubel2004ascending} and its polyhedral extensions~\cite{goel2015polyhedral} implement this for polymatroidal feasible regions.

\textbf{Service-dependency DAGs and sovereignty constraints.}
The AI Service Markets framework of~\cite{trilogy_p1} provides the formal substrate adopted in this paper: AI pipelines are modelled as service-dependency DAGs, polymatroidal feasibility regions arise under tree and series-parallel structures, and integrator encapsulation restores tractable allocation for arbitrary DAGs. We use this framework's polymatroidal allocation, integrator encapsulation, and sovereignty-constraint formulation in our distribution architecture (\cref{sec:DistributionArch}). Multi-level governance composition --- how partial enforcement composes across credibility levels under asymmetric trust, capacity, or data quality, formalised on the supermodular-lattice tradition~\cite{topkis1998supermodularity,milgrom1990rationalizability} --- is the subject of forthcoming companion work; testing such composition predictions empirically requires heterogeneous-domain regimes and is out of scope for the homogeneous-load campaign of this paper. We adopt the sovereignty-as-coordinate-wise-bound formulation directly; multi-level credibility composition is left to follow-on work.

\textbf{Multi-agent orchestration and semantic routing.}
LLM-based multi-agent systems~\cite{guo2024llmmultiagent} have produced frameworks for agentic orchestration (AutoGen~\cite{wu2023autogen}, CrewAI~\cite{crewai2024}, OpenAI Agents SDK~\cite{openai_agents2025}); emerging work on edge-deployed multi-LLM systems (e.g.,~\cite{chen2025edgeintelligence}) addresses trust, dynamic orchestration, and resource scheduling across heterogeneous edge nodes. These systems orchestrate agents (selecting which LLM handles a sub-task), but assume a shared runtime and do not address how agents discover and consume each other's outputs across administrative boundaries. Manias et al.~\cite{manias2024semantic} apply embedding-based semantic routing to intent-based 5G core network management, achieving deterministic intent classification but limited to a fixed set of management actions. Programmable forwarding planes (P4Runtime~\cite{p4runtime}, INT~\cite{kim2015int}) are data-plane orthogonal to the present scope.

\textbf{Gap.}
Existing market mechanisms allocate resources but lack a semantic interconnect layer for cross-domain coordination; cross-domain self-management substrates with provable welfare guarantees and runtime sovereignty enforcement have not been demonstrated in networked deployments. Neural Pub/Sub addresses these gaps: the broker federation integrates semantic matching with autonomic market-based allocation (\cref{sec:market-allocation}), and the sovereignty module enforces tenant- and operator-specified data residency at zero measurable runtime cost (\cref{sec:gov-composition,sec:f4}).

\subsection{Positioning}\label{sec:rw-position}

The Neural Router~\cite{neuralrouter2025} is a single-broker semantic-matching system in the same line of work; the present paper specifies a federated-broker autonomic substrate that adds (1)~an autonomic federated-market substrate grounded in Walrasian equilibrium theory~\cite{kelso1982job,trilogy_p1} and the MAPE-K reference model~\cite{kephart2003vision,ibm2003autonomic}; (2)~runtime sovereignty enforcement at zero measurable cost via the polymatroidal feasibility region; and (3)~a tiered experimental design validating both infrastructure properties and scientific hypotheses. The result is a system that uniquely combines semantic content matching, multi-stage pipeline composition, cross-domain federation, sovereignty enforcement, and market-aware resource allocation under a single MAPE-K loop, distinguished from existing systems that support at most a strict subset of these dimensions (Kafka / Flink / Ray, MLflow / Kubeflow / Argo and Kubernetes-side schedulers cover multi-stage but not cross-domain or semantic-aware allocation; SIENA~\cite{carzaniga2001design} and content-based pub/sub~\cite{tarkoma2012publish} cover cross-domain but not multi-stage or sovereignty; DeepSC~\cite{xie2021deep} and 5G-semantic-routing~\cite{manias2024semantic} cover semantic matching but not pipelines; AI Service Markets~\cite{trilogy_p1} cover multi-stage / sovereignty / market but lack semantic matching; LLM multi-agent~\cite{guo2024llmmultiagent} cover only partial slices). The original Neural Pub/Sub paradigm appeared in~\cite{loven2023can}; \cite{saleh2025messagebrokers,tarkoma2023ai} surveyed broker and AI-native interconnect frameworks. This paper provides the first formal specification of the autonomic distributed substrate that realises the vision, with the first empirical validation of decentralised-vs-centralised parity at equal process count.
           %% §2 Related Work (autonomic-computing canon + orchestration positioning)

\section{System Model}\label{sec:SystemModel}

This section establishes the formal foundations for the distribution architecture of \cref{sec:DistributionArch}. We first introduce service-dependency DAGs from the AI Service Markets framework~\cite{trilogy_p1} (\cref{sec:service-dag}), then state the Walrasian convergence result that motivates the architecture's market mechanism (\cref{sec:walrasian-convergence}), describe the broker's content-matching layer (\cref{sec:neural-router-recap}), and define the design patterns that compose AI pipelines (\cref{sec:design-patterns}). Symbols are introduced inline at first use. The clearing price $\pi_{k,t}$ in \cref{sec:market-allocation} is distinct from the placement function $\pi$ defined in \cref{sec:formal}; the symbol's role is unambiguous from context.

\subsection{Service-Dependency DAGs}\label{sec:service-dag}

\subsubsection{Resource Graph and Feasibility}

We model the computing continuum as a set of service types $\mathcal{R}$, ranging from infrastructure primitives (CPU, GPU, bandwidth) through data-processing and inference endpoints to compound agentic capabilities. Services exhibit structural dependencies: an inference service may depend on a pre-processing pipeline and a model-hosting service, each of which depends on underlying compute and storage. These dependencies are captured by a \emph{service-dependency DAG}
\begin{equation}\label{eq:res-dag}
    G_{\mathrm{res}} = (\mathcal{R}, E),
\end{equation}
where a directed edge $(r, r') \in E$ indicates that service $r'$ depends on service $r$. Each service $v \in \mathcal{R}$ has a capacity $C_v$ representing throughput (e.g., requests per second for an inference endpoint, cores for raw compute).

Agents consume \emph{leaf services} $L(G) \subseteq \mathcal{R}$, the terminal nodes of the DAG that represent externally accessible service endpoints. Each internal node $v$ constrains the aggregate throughput of its descendant leaves: the set of leaves reachable from $v$, denoted $L_v$, satisfies
\begin{equation}\label{eq:res-feasibility}
    \sum_{l \in L_v} x_l \le C_v,
\end{equation}
where $x_l$ is the throughput allocated to leaf $l$. The \emph{service-feasibility region} is the intersection of all such constraints:
\begin{equation}\label{eq:res-region}
    \mathcal{X}_{\mathrm{res}} = \bigl\{ \bm{x} \ge 0 : x(L_v) \le C_v \text{ for every internal node } v \bigr\},
\end{equation}
where $x(S) = \sum_{l \in S} x_l$.

\subsubsection{Polymatroidal Structure}

The shape of $\mathcal{X}_{\mathrm{res}}$ depends on the DAG topology. When $G_{\mathrm{res}}$ is a rooted tree or a two-terminal series-parallel network, the constraint families $\{L_v\}$ form a \emph{laminar family} (any two members are either nested or disjoint), and $\mathcal{X}_{\mathrm{res}}$ is a \emph{polymatroid} with rank function~\cite{trilogy_p1}
\begin{equation}\label{eq:rank}
    f(S) = \min_{A} \sum_{v \in A} C_v,
\end{equation}
minimised over antichains $A$ such that $S \subseteq \bigcup_{v \in A} L_v$. A polymatroid is the polytope $\{\bm{x} \ge 0 : x(S) \le f(S)\; \forall S \subseteq L(G)\}$ for a normalised, monotone, submodular rank function $f$~\cite{fujishige2005submodular}. This structure enables efficient allocation via greedy algorithms or ascending auctions and incentive-compatible pricing under gross-substitutes valuations~\cite{kelso1982job,gul1999grosssubstitutes,trilogy_p1}.

\emph{Why tree/SP DAGs induce laminar constraint families.} A two-terminal SP digraph admits a binary parse tree~\cite{duffin1965topology,valdes1982recognition} whose leaves are $L(G)$; for every internal node $v$, $L_v$ coincides with the descending parse-tree leaves, so $\{L_v\}$ is laminar on $L(G)$ (rooted trees are the single-edge-operand special case). Laminarity makes \cref{eq:rank} monotone, normalised, and submodular, so $\mathcal{X}_{\mathrm{res}}$ is a polymatroid~\cite{edmonds1970submodular,fujishige2005submodular}. General DAGs may destroy laminarity: the canonical obstruction is the diamond ($K_4$-minor), and the \emph{ran-entangled} pipeline (\cref{sec:scenario}) contains such a diamond, resolved by integrator encapsulation contracting each domain's sub-DAG to a single composite node.

\subsubsection{Integrator Encapsulation and Governance Constraints}\label{sec:integrator-concept}

The AI Service Markets framework~\cite{trilogy_p1} resolves general-DAG complexity through \emph{integrator encapsulation}: an integrator manages a connected sub-DAG and exposes a composite service of scalar capacity
\begin{equation}\label{eq:integrator-cap}
    \bar{C}_j = \mathrm{max\text{-}flow}(G_j).
\end{equation}
Contracting each integrator yields a quotient graph $G'$; when $G'$ is tree or SP, the agent-facing feasibility region is polymatroidal. \Cref{sec:DistributionArch} instantiates each domain's broker as an integrator. Governance constraints impose coordinate-wise upper bounds $u_l \le C_l$ on leaf allocations~\cite{trilogy_p1}; the governance-constrained region $\mathcal{X}_t = \mathcal{X}_{\mathrm{res}} \cap \mathcal{X}_{\mathrm{gov}}$ remains polymatroidal (intersection with coordinate-wise bounds preserves polymatroidal structure~\cite{fujishige2005submodular}).

\subsection{Theoretical Anchor: Walrasian Convergence on Tree/SP DAGs}\label{sec:walrasian-convergence}

The distribution architecture's market mechanism (\cref{sec:market-allocation}) is motivated by the following convergence result; the implemented mechanism is a marginal-cost approximation rather than the full polyhedral-clinching auction (gap discussed in \cref{sec:market-impl,sec:f1}). The self-contained statement with conditions is in Section~S2 of the electronic supplement.

\begin{proposition}[Walrasian convergence for tree/SP DAGs]\label{prop:walrasian}
Let $G_{\mathrm{res}}$ have tree or SP structure with laminar $\{L_v\}$ on $L(G)$, and let agents' valuations on leaf-allocation slices satisfy gross-substitutes (GS)~\cite{kelso1982job}. Then a Walrasian equilibrium $(\bm{p}^*,\bm{x}^*)$ exists, $\bm{x}^*$ maximises social welfare, and the equilibrium is computable in polynomial time via an ascending polyhedral-clinching auction on the polymatroidal feasibility region~\cite{ausubel2004ascending,goel2015polyhedral}.
\end{proposition}

\noindent The result combines tree/SP polymatroidal structure~\cite{fujishige2005submodular,trilogy_p1} with Kelso--Crawford~\cite{kelso1982job} and Gul--Stacchetti~\cite{gul1999grosssubstitutes}. The GS condition is non-trivial: raw pipeline-bundle valuations are Leontief; the GS-compatible representation arises only after integrator encapsulation (\cref{sec:integrator-concept}) bundles multi-resource paths into composite slices with unit demand.

\subsection{The Broker's Content-Matching Layer}\label{sec:neural-router-recap}\label{sec:semcomm-positioning}

Each broker performs content matching by embedding subscriptions and incoming events into a dense vector space and assigning each event to subscriptions whose embeddings exceed a similarity threshold $\tau$ within $k$-means clusters. The single-broker matching layer's design, calibration of $\tau$, embedding model, and optional LLM-based cluster compression are companion-paper concerns (\cite{neuralrouter2025}); \emph{the present paper exercises the matching layer as a frozen, pre-calibrated component} so that the experimental signal is attributable to the distribution architecture (\cref{sec:DistributionArch}) rather than to matching-quality variation. Subscription cluster summaries $(\bar{\bm{e}}_{k,i},r_{k,i})$ in \cref{eq:summary} are rate-distortion-style summaries of the underlying subscription set, complementary to semantic-communication transport codecs~\cite{xie2021deep,bourtsoulatze2019deep,strinati2021semantic} (the matching layer operates one layer up, on already-decoded content, deciding which receivers should act on it). Live integration with dynamic agent runtimes (LangGraph~\cite{langgraph2024}, AutoGen~\cite{wu2023autogen}, Anthropic MCP~\cite{anthropic_mcp2024}, Google A2A~\cite{google_a2a2025}) -- where the matching layer would route agent requests rather than pre-registered subscriptions -- is deferred and is discussed as future work (\cref{sec:Discussion}).

\subsubsection{Design Patterns and the Pipeline Placement Problem}\label{sec:design-patterns}

Multi-stage AI workloads compose two design patterns into pipelines $G_{\text{pipe}}=(V,E)$: a \emph{map} operation ($1\to F\to 1$) transforms a publisher's event through a processing function $F$ before delivery (e.g., classification, summarisation), and a \emph{funnel} ($N\to F\to 1$) aggregates events from multiple publishers (e.g., sensor fusion, federated aggregation). These patterns are analogous to MapReduce~\cite{dean2008mapreduce} but differ in two respects: the broker dynamically determines participation through semantic matching, and the patterns operate in streaming mode across the computing continuum. The polymatroidal feasibility argument of \cref{sec:service-dag} subsumes both patterns: the parse-tree argument applies whenever the composed pipeline is series-parallel after integrator encapsulation. Given $G_{\text{pipe}}$ and execution units distributed across multiple administrative domains, the \emph{pipeline placement problem} finds a mapping $\pi: V \to \mathcal{N}$ that assigns each stage to a node, subject to capacity, latency, and governance constraints, minimising a weighted combination of total inter-stage latency, resource utilisation imbalance, and administrative domain crossings; \cref{sec:DistributionArch} formalises the constraints and presents the market-based, oracle, and heuristic placement algorithms.

\subsection{Edge-Cloud Continuum Topology}\label{sec:oran-mapping}

The abstract multi-domain model of \cref{sec:service-dag,sec:design-patterns} is instantiated as a four-domain edge-cloud topology spanning two sites:
\begin{itemize}
    \item \textbf{Edge site:} Domains $d_1$ and $d_2$. Low-latency compute co-located with edge devices. Within-site latency $< 1$\,ms.
    \item \textbf{Cloud site:} Domains $d_3$ and $d_4$. Abundant compute at higher latency. Within-site latency $< 1$\,ms.
\end{itemize}
A single WAN link ($\sim$50\,ms) connects the two sites, representing the midhaul/backhaul boundary of the continuum. Each domain $d_k\in\mathcal{D}$ is administratively independent, with its own governance policy $\mathcal{G}_k$, compute pool $\mathcal{N}_k$, and trust relationships (the formal multi-domain model is given in \cref{sec:formal}). The topology creates the central autonomic placement question: should a pipeline stage execute at the edge (low latency, constrained compute) or in the cloud (high latency, abundant compute)? The price signals of \cref{sec:market-allocation} resolve this trade-off without global state. \Cref{sec:scenario} maps the four logical domains onto a concrete O-RAN deployment for the evaluation, but the substrate is O-RAN-agnostic; any four-administrative-domain edge-cloud topology suffices.

The evaluation exercises three representative 8-stage pipeline structures that span the polymatroidal-tractability spectrum, parameterised by the \emph{non-modularity gap} $\gamma$ (the deficit of a constraint family from being laminar; $\gamma=0$ for tree/SP DAGs, $\gamma>0$ when the rank function fails submodularity, defined formally in \cref{sec:scenario}): a \emph{linear chain} (tree, $\gamma = 0$), where Walrasian convergence holds directly; a \emph{fan-in series-parallel} pattern (SP, $\gamma = 0$), where convergence holds via the parse-tree argument of \cref{sec:service-dag}; and an \emph{entangled} cross-tree fan-out/fan-in pattern ($\gamma > 0$), where convergence holds only after integrator encapsulation contracts each domain to a composite node. The concrete instantiation of these as O-RAN cross-layer pipelines (cqi-chain, anomaly-sp, ran-entangled) is given in \cref{sec:scenario}.
           %% §3 System Model (service-DAG, propositions, content-matching layer)

\section{Distribution Architecture}\label{sec:DistributionArch}

The single-broker content-matching layer of \cref{sec:neural-router-recap} performs learned semantic matching within one administrative domain. The computing continuum spans multiple administrative domains, network slices, and geographic sites; closing an autonomic control loop over this continuum requires federating multiple broker instances into a coherent self-managing fabric. This section specifies the \emph{distribution architecture}, including the autonomic-market allocation mechanism (\cref{sec:market-allocation}) that implements the Walrasian convergence prediction of \cref{sec:walrasian-convergence}, the sovereignty-enforcement scenarios (\cref{sec:gov-composition}), and the explicit MAPE-K mapping (\cref{sec:mapek}) that frames the substrate as an autonomic computing system in the sense of~\cite{kephart2003vision,ibm2003autonomic,salehie2009self,parashar2025everywhere}.

\subsection{Formal Model}\label{sec:formal}

\subsubsection{Domains, Brokers, and Execution Units}

Let $\mathcal{D} = \{d_1, \ldots, d_m\}$ be a set of \emph{administrative domains}, each governed by an independent operator. Each domain $d_k$ contains:
\begin{itemize}
    \item A \emph{neural broker} $b_k$ implementing the content-matching layer of \cref{sec:neural-router-recap};
    \item A set of \emph{execution units} $\mathcal{N}_k = \{n_{k,1}, \ldots, n_{k,|\mathcal{N}_k|}\}$, each with computational capacity $C_{k,j}$;
    \item A set of locally registered publishers $\mathcal{P}_k$ and subscribers $\mathcal{S}_k$.
\end{itemize}

\subsubsection{AI Pipelines as Service-Dependency DAGs}

Following the AI Service Markets framework~\cite{trilogy_p1}, an AI pipeline is modelled as a \emph{service-dependency DAG} $G_{\text{pipe}} = (V, E)$, where each node $v \in V$ represents a processing stage and each directed edge $(v, v') \in E$ indicates that stage $v'$ depends on the output of stage $v$. Each stage $v$ has a \emph{computational demand} $\rho_v$ and an \emph{output data rate} $\omega_v$. A pipeline \emph{placement} is a mapping $\pi: V \to \bigcup_k \mathcal{N}_k$ that assigns each stage to an execution unit. A placement is \emph{feasible} if:
\begin{equation}\label{eq:capacity}
    \sum_{v : \pi(v) = n_{k,j}} \rho_v \le C_{k,j} \quad \forall k, j
\end{equation}
and, for every edge $(v, v') \in E$:
\begin{equation}\label{eq:latency}
    \ell(\pi(v), \pi(v')) \le L_{v,v'}
\end{equation}
where $\ell(\cdot,\cdot)$ denotes the network latency and $L_{v,v'}$ is the latency bound.

\subsubsection{Governance Constraints}

A \emph{governance policy} $\mathcal{G}_k$ for domain $d_k$ specifies:
\begin{itemize}
    \item \textbf{Data sovereignty:} a set $\mathcal{T}_k^{\text{local}} \subseteq V$ of stage types whose inputs must not leave domain $d_k$;
    \item \textbf{Trust requirements:} for cross-domain edges, a minimum trust level $\tau_{k,k'}$ between domains;
    \item \textbf{Audit obligations:} stages processing regulated data must produce verifiable evidence bundles.
\end{itemize}
A placement $\pi$ is \emph{governance-compliant} if it satisfies \cref{eq:capacity,eq:latency} and:
\begin{equation}\label{eq:governance}
    v \in \mathcal{T}_k^{\text{local}} \implies \pi(v) \in \mathcal{N}_k
\end{equation}

\subsection{Broker Federation}\label{sec:federation}

\subsubsection{Federation Overlay}

The brokers form a \emph{federation overlay} $\mathcal{F} = (\mathcal{B}, E_F)$ where $\mathcal{B} = \{b_1, \ldots, b_m\}$ and $(b_k, b_{k'}) \in E_F$ if domains $d_k$ and $d_{k'}$ have established a federation agreement (mutual trust $\tau_{k,k'} > 0$).

\subsubsection{Subscription Aggregation and Propagation}\label{sec:aggregation}

Each broker $b_k$ maintains a local subscription index: a set of clusters $\{c_{k,1}, \ldots, c_{k,p_k}\}$ with associated embeddings and optimised subscription prompts. To enable cross-domain routing, brokers exchange \emph{aggregated subscription summaries} with federation peers.

\begin{definition}[Subscription Summary]
The subscription summary of broker $b_k$ is:
\begin{equation}\label{eq:summary}
    \sigma_k = \left\{ \left( \bar{\bm{e}}_{k,i}, \; r_{k,i}, \; \kappa_{k,i} \right) : i = 1, \ldots, p_k \right\}
\end{equation}
where $\bar{\bm{e}}_{k,i}$ is the centroid embedding of cluster $c_{k,i}$, $r_{k,i}$ is the cluster radius, and $\kappa_{k,i}$ is the available capacity.
\end{definition}

\subsubsection{Cross-Domain Routing, Consistency, and Recovery}\label{sec:consistency}\label{sec:routing}

When a publication arrives at broker $b_k$ and the local content-matching layer finds no adequate match, the broker initiates \emph{federated routing}: embed publication, select peer-cluster pairs within distance threshold, apply governance filter, forward to selected brokers ranked by semantic distance, aggregate responses by confidence score; $O(m\cdot\bar{p})$ comparison cost per publication using precomputed centroids. The overlay is asynchronous: peers exchange price summaries every $\delta_{\text{prop}}=10$\,s, with $\text{WAN}_{\max}\approx 50$\,ms giving a bounded staleness $B\approx 10.05$\,s; routing decisions via \cref{eq:trade} use stale-but-bounded prices, corrected by the next exchange. Each federated dispatch carries a per-RPC timeout $\tau_{\text{fed}}=5$\,s; on miss, the originating broker treats the peer as unreachable and falls back to local-only placement. The same $\delta_{\text{prop}}$-period channel doubles as the broker-liveness signal: after $k_{\text{miss}}=3$ consecutive failed pushes ($\approx 30$\,s) the peer is marked unhealthy, its $\sigma_k$ and price signal evicted, and it is excluded from routing until a recovery probe (every five rounds) succeeds; on partition healing, brokers exchange the last $B$\,s of price history and resume periodic exchange (no global rollback).

\paragraph{Consistency model and delivery semantics.} The federation overlay's Knowledge layer (peer subscription summaries $\sigma_k$, peer price signals; \cref{sec:mapek}) is \emph{bounded-staleness eventually consistent} in the sense of probabilistically-bounded staleness~\cite{bailis2014pbs}: any read at broker $b_k$ reflects a peer state at most $B$ seconds old, and across-peer convergence is guaranteed within one $\delta_{\text{prop}}$ period after the network heals. Subscription propagation and price exchange are POST-only HTTP RPCs with at-most-once delivery semantics (no retries on the application side) and no global ordering guarantee; the orchestration result is therefore conditional on the assumption that loss / reorder are below the levels at which the $k_{\text{miss}}=3$ failover regime begins to misclassify peers. Pipeline dispatch from broker to local worker is at-least-once with idempotent stage handlers (worker-side deduplication on pipeline ID), and the within-round ledger $\ell_{\text{add}}$ (\cref{sec:market-impl}) is single-writer per broker and committed atomically at end-of-round. \emph{Cross-broker reservation conflicts cannot occur} by construction: each worker is registered with exactly one broker, so two brokers cannot dispatch to the same worker concurrently; cross-broker over-allocation onto a domain (rather than a worker) is bounded by a single pipeline's stage count and settled within $B+\delta_{\text{health}}$, consistent with the asynchronous, eventually-consistent semantics of \cref{prop:walrasian}.

\subsubsection{Integrator Encapsulation}\label{sec:integrator}

Each domain's broker acts as an \emph{integrator}~\cite{trilogy_p1}: it encapsulates the domain's internal service-dependency structure into a composite service with scalar capacity:
\begin{equation}\label{eq:composite}
    \hat{C}_k^t = \max\text{-flow}\left(G_k^t\right)
\end{equation}
The federation-level allocation operates on the quotient graph $\hat{G} = (\hat{\mathcal{D}}, \hat{E})$ where each domain is a single node. When $\hat{G}$ is tree or series-parallel, the federation-level feasibility region is polymatroidal (\cref{prop:walrasian}), and efficient mechanisms apply.

\subsection{Market-Based Decentralised Allocation}\label{sec:market-allocation}

The market mechanism enables pipeline placement across domain boundaries using price signals rather than global state. This operationalises the Walrasian convergence result of \cref{prop:walrasian}: for tree/SP pipeline DAGs, the price-based mechanism converges to the welfare-maximising allocation.

\subsubsection{Bids, Clearing Prices, Federation, and Trade Decision}

Each worker $n_{k,j}$ in domain $d_k$ submits a bid $\text{WorkerBid}(n_{k,j}, t) = c_{k,j,t}$ for each stage type $t$, with $c_{k,j,t}$ the opportunity cost increasing in utilisation (congestion pricing). The domain broker $b_k$ computes per-stage-type clearing prices via marginal cost pricing,
\begin{equation}\label{eq:clearing}
    \pi_{k,t} = c_{k,(d_t),t},
\end{equation}
where $c_{k,(d_t),t}$ is the $d_t$-th cheapest worker's cost for stage type $t$ and $d_t = \min(\text{demand}_t, \text{supply}_{k,t})$. Brokers exchange price signals
\begin{equation}\label{eq:pricesignal}
    \text{PriceSignal}(d_k) = \{(t,\pi_{k,t}) : t\in\text{stage types}\}
\end{equation}
alongside subscription summaries (preserving worker-identity privacy). For each stage of type $t$, the originating broker compares local and remote prices:
\begin{equation}\label{eq:trade}
    \text{trade}(t) = \begin{cases}
        \text{remote} & \text{if } \pi_{k',t} + w_{\text{WAN}} < \pi_{k,t}, \\
        \text{local}  & \text{otherwise},
    \end{cases}
\end{equation}
with $w_{\text{WAN}}$ the WAN transfer cost; ties prefer local placement, and the pipeline is accepted if total cost $\le$ pipeline value budget. \Cref{alg:market} summarises the procedure (runs per allocation epoch).

\begin{algorithm}[t]
\caption{Market-Based Pipeline Placement}
\label{alg:market}
\begin{algorithmic}[1]
\Require Pipeline $G_{\text{pipe}} = (V, E)$, local domain $d_k$, worker bids, price signals from federation peers, WAN cost $w_{\text{WAN}}$
\Ensure Placement $\pi: V \to \mathcal{N}$ or rejection
\Statex \textit{Concurrency:} round-synchronous; bids are collected before clearing, peer prices are read-only within a round, and pipelines are processed in arrival order at each broker.
\State Collect WorkerBids from all workers in $d_k$
\State Compute clearing prices $\pi_{k,t}$ for each stage type $t$ via marginal cost pricing (\cref{eq:clearing}); equal-cost bids ordered by ascending worker ID
\State Broadcast $\text{PriceSignal}(d_k)$ to federation peers
\For{each stage $v \in V$ in DAG topological order (Kahn's, ties by ascending stage ID)}
    \State $t \gets \text{type}(v)$
    \For{each remote domain $d_{k'}$ with price signal}
        \If{$\pi_{k',t} + w_{\text{WAN}} < \pi_{k,t}$}
            \State Mark $v$ for remote placement in $d_{k'}$
        \EndIf
    \EndFor
    \State Assign $v$ to $\arg\min$ over feasible workers of cost-per-stage in the chosen domain (ties broken by ascending worker ID)
    \State Accumulate cost
\EndFor
\If{total cost $>$ pipeline value budget}
    \State \Return rejection
\EndIf
\State \Return placement $\pi$
\end{algorithmic}
\end{algorithm}

The clearing prices $\pi_{k,t}$ approximate Walrasian prices: they reflect the marginal cost of the scarce resource. For tree/SP DAGs where workers exhibit GS valuations (single stage type, unit demand), the price adjustment converges to the equilibrium prices of \cref{prop:walrasian}; \cref{sec:Evaluation} tests how closely this practical mechanism matches the centralised oracle.

\subsubsection{Implementation refinements}\label{sec:market-impl}

The Walrasian mechanism in \cref{prop:walrasian} is realised through five coupled components, each addressing a distinct gap that emerged during testbed validation: \emph{(i) speed-scaled bids} $b_i = b_0\,s_i$ at registration encode intrinsic worker capability ($s_i$ = relative processing speed) into the price signal independent of load (utilisation-based pricing alone produces only $\sim 2$\,ms price gaps at $\lambda=5$\,pps, below any realistic WAN cost); \emph{(ii) M/M/1 dynamic congestion pricing} computes per-worker cost
\begin{equation}\label{eq:mm1-cost}
\text{cost}_i = \frac{b_i}{1 - \rho_i},\quad \rho_i = \min\!\left(\frac{\ell_i}{c_i},\, 0.99\right),
\end{equation}
with the $\rho_i\le 0.99$ cap inside the formula bounding $\text{cost}_{\max}=100\,b_i$ (orders of magnitude above $w_{\text{WAN}}$, so WAN-vs-local trade is dominated by genuine queue-depth signal); \emph{(iii) federated price exchange} (POST \texttt{/federation/price-signal} every $\delta_\text{prop}$\,s, alongside subscription-summary propagation) ensures routing decisions reflect federation-wide scarcity rather than only the local domain's clearing price; \emph{(iv) within-round load reservation} via a per-round commitment ledger $\ell_\text{add}: \mathcal{U}\to\mathbb{R}_{\ge 0}$ tests $\rho_v\le c_u - \ell_u - \ell_\text{add}(u)$ rather than the pre-round residual snapshot, preventing an 8-stage cqi-chain pipeline from over-committing a single cheap worker; \emph{(v) health-check utilisation synchronisation} (one probe per worker per $\delta_\text{health}=5$\,s) keeps each broker's view of $\ell_i$ — the input to \cref{eq:mm1-cost} — responsive to actual worker queue depth rather than a stale local estimate. The five are jointly sufficient for the empirical envelope tested in \cref{sec:Evaluation}: each component addresses a concrete failure mode encountered during testbed validation, and the campaign reported here exercises the substrate with all five engaged. A per-component leave-one-out attribution table — which would isolate the marginal contribution of each refinement — is deferred to follow-on work; the campaign's three-property Walrasian decomposition (information completeness, admission control, price discovery; \cref{sec:f2}) instead isolates which Walrasian \emph{properties} are exercised, not which engineering refinements are individually load-bearing.

\subsection{Autonomic Control Loop (MAPE-K)}\label{sec:mapek}

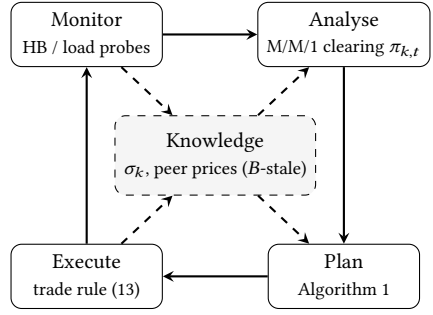
\begin{figure}[t]
    \centering
    \begin{tikzpicture}[
        box/.style={draw, rounded corners=4pt, minimum height=0.85cm, minimum width=2.0cm, align=center, font=\footnotesize},
        kbox/.style={draw, dashed, rounded corners=4pt, minimum height=1.05cm, minimum width=2.2cm, align=center, font=\footnotesize, fill=gray!8},
        arr/.style={->, >=stealth, thick}
    ]
    \node[box] (M) at (0, 1.6) {Monitor\\\scriptsize HB / load probes};
    \node[box] (A) at (3.4, 1.6) {Analyse\\\scriptsize M/M/1 clearing $\pi_{k,t}$};
    \node[box] (E) at (0, -1.6) {Execute\\\scriptsize trade rule (\ref{eq:trade})};
    \node[box] (P) at (3.4, -1.6) {Plan\\\scriptsize Algorithm~\ref{alg:market}};
    \node[kbox] (K) at (1.7, 0) {Knowledge\\\scriptsize $\sigma_k$, peer prices ($B$-stale)};
    \draw[arr] (M) -- (A);
    \draw[arr] (A) -- (P);
    \draw[arr] (P) -- (E);
    \draw[arr] (E) -- (M);
    \draw[arr, dashed] (M) -- (K);
    \draw[arr, dashed] (K) -- (A);
    \draw[arr, dashed] (K) -- (P);
    \draw[arr, dashed] (E) -- (K);
    \end{tikzpicture}
    \caption{MAPE-K control loop closed at each broker. Solid arrows: control flow per allocation epoch. Dashed arrows: Knowledge updates and reads. Knowledge is bounded-staleness (\cref{sec:consistency}, $B\approx 10.05$\,s) rather than synchronously consistent.}
    \label{fig:mapek}
\end{figure}

The components of \cref{sec:formal,sec:federation,sec:market-allocation} jointly close a MAPE-K control loop~\cite{kephart2003vision,ibm2003autonomic} at each broker, framing Neural Pub/Sub as an autonomic substrate in the sense of self-adaptive software~\cite{salehie2009self,cheng2009software,weyns2020software} (\cref{fig:mapek}). The five elements are realised concretely as follows.

\textbf{Monitor.} Each broker tracks the health and load of its local workers via the periodic probe loop ($\delta_{\text{health}}=5$\,s, \cref{sec:market-impl}) and tracks federation peers via the heartbeat-based partition detector ($\delta_{\text{prop}}=10$\,s exchange period, $k_{\text{miss}}=3$ consecutive misses; \cref{sec:consistency}). The monitor produces per-worker queue depth $\ell_i$ (input to~\cref{eq:mm1-cost}) and per-peer liveness, both with bounded latency.

\textbf{Analyse.} The broker computes per-stage-type clearing prices $\pi_{k,t}$ from worker bids using the M/M/1 sojourn-time cost model (\cref{eq:mm1-cost,eq:clearing}). This is the marginal-cost analysis stage of the autonomic loop: it converts raw monitored state into decision-relevant signals, with the $\rho_i\le 0.99$ cap providing bounded admission control. Speed-scaled bids (\cref{sec:market-impl}) ensure the analysis differentiates intrinsic worker capability from current load.

\textbf{Plan.} Algorithm~\ref{alg:market} computes a placement $\pi: V\to\bigcup_k\mathcal{N}_k$ over the polymatroidal feasibility region $\mathcal{X}_t = \mathcal{X}_{\mathrm{res}}\cap\mathcal{X}_{\mathrm{gov}}$ (\cref{sec:service-dag}). Planning is local to each broker and uses cached peer prices for cross-domain trade decisions (\cref{eq:trade}). \Cref{prop:walrasian} guarantees that, under tree/SP DAG structure and gross-substitutes valuations, the planned allocation converges to the welfare-maximising Walrasian equilibrium; the marginal-cost approximation gap is characterised in \cref{sec:f1}.

\textbf{Execute.} The planned placement is dispatched as federated cross-domain HTTP calls (per-RPC timeout $\tau_{\text{fed}}=5$\,s, \cref{sec:consistency}), with the per-round reservation ledger $\ell_{\text{add}}$ committing atomically at end-of-round (\cref{sec:market-impl}). Execution updates the local Monitor state (worker load reports) and pushes a fresh price-signal vector (POST \texttt{/federation/price-signal}) to peers, closing the loop.

\textbf{Knowledge.} The shared Knowledge layer holds two artefacts: per-peer subscription summaries $\sigma_k = \{(\bar{\bm{e}}_{k,i}, r_{k,i}, \kappa_{k,i})\}$ (\cref{eq:summary}) for cross-domain semantic routing, and cached peer price signals $\text{PriceSignal}(d_{k'})$ (\cref{eq:pricesignal}) for cross-domain trade decisions. Both are bounded-staleness rather than synchronously consistent: the staleness bound $B = \delta_{\text{prop}} + \text{WAN}_{\max} \approx 10.05$\,s (\cref{sec:consistency}) is the load-bearing operational parameter that decouples the autonomic loop from synchronous federation handshakes. The within-round reservation ledger and the next-round price exchange jointly absorb the staleness, consistent with the eventually-consistent semantics of \cref{prop:walrasian}.

The MAPE-K mapping makes three engineering claims explicit. First, the loop is \emph{decentralised}: every broker closes its own loop, with no global coordinator on the critical path. Second, the loop is \emph{economically grounded}: the Analyse and Plan stages use mechanism-design primitives (marginal-cost clearing, polymatroidal feasibility, Walrasian equilibrium) rather than rule-based heuristics or feedback-control gain tuning, distinguishing this work from the bulk of self-adaptive substrates. Third, the loop is \emph{bounded-staleness} rather than strongly consistent: the Knowledge layer's $B$-second staleness is a designed parameter, not a defect, and the autonomic robustness results of \cref{sec:f3} directly test loop behaviour under broker death and network partition.

\subsection{Tiered Placement Capabilities and Slice-Aware Cost}\label{sec:placement}\label{sec:slice-aware}

The architecture supports a hierarchy of placement capabilities for controlled experimental comparison: \emph{oracle} (single broker, full global visibility, congestion-aware DP/greedy on \cref{eq:cost}; efficiency upper bound), \emph{conventional centralised round-robin} (single broker, full visibility but no cost-optimised placement; conventional-orchestrator baseline), \emph{market} (four brokers, price-signal coordination via \cref{alg:market}; \cref{prop:walrasian}'s mechanism), and three \emph{heuristics} (locality-only, latency-greedy, spillover). All five honour slice constraints $\pi(v)\in\mathcal{N}^{s(v)}$ as a feasibility filter (5G/6G network slices~\cite{tgpp23501,tgpp23502,tgpp28530,3gpp_nrm} partition the physical infrastructure; we model two QoS tiers, low-latency 1\,ms and best-effort 5\,ms added processing delay, labelled URLLC/eMBB in the evaluation). The placement cost function is
\begin{multline}\label{eq:cost}
    \pi^* = \arg\min_{\pi} \biggl[
      \alpha \sum_{(v,v') \in E} \ell\bigl(\pi(v), \pi(v')\bigr)
      + \beta \sum_{v \in V} \frac{\rho_v}{C_{\pi(v)}}   
      + \zeta \bigl|\{k : \exists v,\, \pi(v) \in \mathcal{N}_k\}\bigr|
    \biggr]
\end{multline}
with three terms for inter-stage latency, per-stage resource utilisation, and domain crossings. The placement-cost weight $\zeta$ is distinct from the non-modularity gap $\gamma$ of \cref{sec:scenario}. Tree-structured pipelines admit a $O(|V|\cdot|\mathcal{N}|^2 + |\mathcal{B}|\log W)$ DP solution where $|\mathcal{B}|\le m\cdot|\mathcal{T}|\cdot W$ is the bid-space cardinality and $W=\max_k|\mathcal{N}_k|$; general DAGs use greedy topological assignment in $O(|V|\cdot|\mathcal{N}|+|\mathcal{B}|\log W)$. When $\zeta>0$, the domain-crossings term breaks tree decomposition by a $2^{|\mathcal{D}|}$ factor; with $|\mathcal{D}|=4$ in the deployment this is bounded by 16 and dominated by the polynomial term at the testbed scale ($|V|=8$, $|\mathcal{N}|=48$).

\subsection{Sovereignty Enforcement Scenarios}\label{sec:gov-composition}

The sovereignty module implements four enforcement scenarios at the site level (edge $\{d_1,d_2\}$ and cloud $\{d_3,d_4\}$ each either enforce or do not): \emph{Scenario A} (neither enforces), \emph{B} (edge-only), \emph{C} (cloud-only, symmetric to B), \emph{D} (both enforce). The 2$\times$2 design exercises sovereignty enforcement at increasing site-level coverage and is used in \cref{sec:f4} to verify that enforcement adds no measurable runtime cost across the four regimes.

\subsection{Failure Handling and Scaling}\label{sec:failures}\label{sec:scaling}

If a broker fails, federation peers detect via heartbeat timeouts and continue routing using cached summaries; publishers/subscribers in the failed domain re-register with the nearest peer. When an execution unit fails, the local broker detects via health-check and triggers stage re-placement to an alternative node. The funnel pattern supports three failure modes: wait (buffer and timeout), proceed (partial inputs), or abort (signal failure downstream); per-funnel choice is part of the pipeline's fault-tolerance policy~\cite{carbone2017flink}. For large-scale deployments ($m>50$ domains), hierarchical federation reduces propagation cost from $O(m^2)$ to $O(m)$; summary compression via super-clustering further reduces overhead, and the architecture's overhead (summary propagation, cross-domain routing, placement optimisation) does not require LLM invocations.

      %% §4 Distribution Architecture (federation + market + MAPE-K mapping)

\section{Evaluation}\label{sec:Evaluation}

We evaluate Neural Pub/Sub on a 4-VM, 4-domain, 48-worker testbed deployed on the 5G Test Network Finland (5GTNF). The evaluation is organised around four headline findings introduced in \cref{sec:methodology-overview}, supported by an infrastructure-validity check that establishes the experimental substrate as sound. Per-cell detail is in the electronic supplement.

\subsection{Scenario: O-RAN Cross-Layer Pipelines on the Edge-Cloud Continuum}\label{sec:scenario}

We instantiate the four edge-cloud-continuum domains as DU / CU+near-RT RIC / non-RT RIC / SMO from the O-RAN architecture~\cite{oranwg1oad}; the substrate is O-RAN-agnostic, and any four-administrative-domain edge-cloud topology suffices. In multi-vendor deployments these components may be supplied by different vendors, creating genuine administrative boundaries with distinct governance policies, compute resources, and QoS characteristics; in 5G integrations, the standardised NWDAF service-based interfaces~\cite{3gpp_nwdaf,tgpp23288} (\texttt{Nnwdaf\_AnalyticsInfo}, \texttt{EventsSubscription}, \texttt{MLModelProvision}) are the natural substrate above which Neural Pub/Sub composes its cross-domain placement layer.

We map the four logical domains to a two-site topology representing the edge-cloud continuum:

\begin{itemize}
    \item \textbf{Edge site} (2~VMs): Domain~$d_1$ (DU, 12~URLLC workers) and Domain~$d_2$ (CU + near-RT RIC, 12~mixed URLLC/eMBB workers). Represents cell-site and edge compute.
    \item \textbf{Cloud site} (2~VMs): Domain~$d_3$ (non-RT RIC, 12~eMBB workers) and Domain~$d_4$ (SMO, 12~best-effort workers). Represents regional and central cloud.
\end{itemize}

\noindent Within each site, brokers federate over LAN ($<$1\,ms). Between sites, a single WAN link ($\sim$50\,ms, emulated via \texttt{tc qdisc netem}) represents the midhaul/backhaul boundary. This creates the placement question on which the architecture is judged: should a pipeline stage execute at the edge (low latency, constrained compute) or in the cloud (high latency, abundant compute)?

Three 8-stage pipeline types arise naturally from O-RAN cross-layer optimisation use cases, each with identical stage count but different DAG structures:

\begin{enumerate}
    \item \textbf{CQI Prediction Chain} (tree, $\gamma = 0$): $\text{DU:raw\_cqi} \to \text{DU:denoise} \to \text{CU:normalise} \to \text{CU:feature\_extract} \to \text{RIC:predict} \to \text{RIC:validate} \to \text{nRT:aggregate} \to \text{SMO:report}$. Linear chain crossing all 4~domains. 8~stages, 7~edges.
    \item \textbf{RAN Anomaly Detection} (series-parallel, $\gamma = 0$): Four parallel sources (2~DU + 2~CU) converge at the near-RT RIC for fusion, classification, alerting, and logging. 8~stages, fan-in only.
    \item \textbf{RAN Intelligence Suite} (entangled, $\gamma > 0$): Cross-domain fan-out from CU:feature\_extract to both RIC:cqi\_predict and RIC:anomaly\_detect, with cross-tree fan-in from RIC + non-RT RIC at SMO:handover\_optimise. 8~stages, 10~edges. Shared stages and diamonds create a non-zero non-modularity gap~$\gamma$.
\end{enumerate}

\noindent The three pipeline templates have identical per-stage compute demand $\rho_v$ across templates (each stage type carries the same configured processing time), so the latency differences reported in \cref{sec:f1,sec:f2} are attributable to DAG structure (chain depth, fan-out, cross-tree dependencies) rather than to per-stage compute heterogeneity.

\subsection{Testbed and Configuration}\label{sec:testbed}

The campaign comprises an infrastructure-validity check on a single 5GTNF virtual machine (University of Oulu; 4~CPU cores, 8\,GB RAM, Docker~CE 29.3 on Ubuntu~24.04) with 2~domains and 5~workers, and a main 4-VM campaign with one VM per O-RAN domain (VM1=DU/URLLC; VM2=CU+near-RT RIC/URLLC+eMBB; VM3=non-RT RIC/eMBB; VM4=SMO/best-effort), 12 workers per VM (48 total), 4 brokers, identical hardware (4 cores / 8\,GB / Ubuntu~24.04). Within-site latency: $<$1\,ms (shared LAN). Cross-site WAN: 50\,ms delay + 5\,ms jitter (emulated via \texttt{tc qdisc netem} between VM2 and VM3). Slice QoS emulated per worker group: URLLC at 1\,ms, eMBB at 5\,ms added processing delay. Each VM runs one Docker Compose stack with its broker and $\sim$12~workers using \texttt{network\_mode: host}; brokers federate via the summary propagation protocol. Placement-cost weights $(\alpha,\beta,\zeta)$ in \cref{eq:cost} are set to $(1,1,1)$ in all reported runs; the substrate accepts arbitrary weights at deployment time, and operator-specific tuning is left to deployment, not load-bearing for the present claims.

Three 8-stage O-RAN pipeline templates are used (\cref{sec:scenario}). Poisson arrivals with rate $\lambda \in \{2, 5, 10\}$~pps for the main allocation grid, distributed across domains (each domain's workload generator targets its local broker). Each configuration runs with five independent seeds. Warmup: 4~minutes (increased for 48-worker steady state). Measurement window: 10~minutes. The ablation programme uses additional rates $\lambda \in \{5, 8, 10, 15, 50\}$~pps to sweep across and beyond the empirical saturation knee at ${\sim}13.8$~pps (calibrated via a dedicated sweep, \texttt{scripts/calibrate\_saturation.py}).

The saturation knee (${\sim}13.8$~pps, 95\% CI $[12.5,14.6]$ via bootstrap on a piecewise-linear segmented regression of mean CR vs $\lambda$) is calibrated by sweeping $\lambda\in\{5,\ldots,16,18,20,25,30,40,50\}$~pps under round-robin placement; the knee is a round-robin property, and the market's admission control pushes its own knee beyond 50~pps (\cref{sec:f2}).

\subsection{Methodology Overview and Findings Summary}\label{sec:methodology-overview}

The 4-VM campaign comprises 1005~runs across seven phases (${\sim}129$~hours of cluster wall clock): a 450-run ablation programme isolating Walrasian mechanism properties under saturation, heterogeneity, and worker failure; a 40-cell baseline phase reproducing infrastructure-validity at 4-VM scale; a 50-cell slicing phase exercising governance and slice variants; a 50-cell worker-failure resilience phase; a 60-cell stress phase combining load and failure injection; a 330-cell main market phase comprising the 270-cell allocation grid (six strategies $\times$ three pipelines $\times$ three loads $\times$ five seeds) and the 60-cell governance grid (four enforcement scenarios $\times$ three pipelines $\times$ five seeds); and a 25-cell federation phase (broker kill, network partition, governance, neural, static). A prior single-VM Tier-1 validation (transport orthogonality, slice awareness, governance overhead, failure resilience) on a smaller substrate supports the infrastructure check; we summarise it inline in \cref{sec:infra-results} and treat the 4-VM baseline as its production-substrate reproduction. The campaign maps to four headline findings; per-cell detail is in the electronic supplement.

\begin{enumerate}
    \item Autonomic decentralisation matches centralised at equal process count and dominates on linear-chain pipelines. The federated market beats the single-process oracle by 2--4\% in every (pipeline, load) cell (45 of 45 per-seed wins, sign-test $p\approx 2.8\times 10^{-14}$, Hodges--Lehmann median $-39.6$\,ms, 95\% bootstrap CI $[-44.1,-35.3]$\,ms); a fair-process-count sharded-oracle comparator (4 coordinator processes) keeps the gap within $\pm 1.5\%$ across all 9 cells, with the market dominating on cqi-chain (3/3) and trading within $\pm 5$\,ms on anomaly-sp and ran-entangled (\cref{sec:f1}).

    \item Strategy choice differentiates only under stress. Under uniform load, market $\approx$ locality $\approx$ latency-greedy $\approx$ spillover within ${\sim}5$\,ms; under saturation, heterogeneity, or worker failure at saturating load, only the market preserves performance. The advantage decomposes into three Walrasian properties: information completeness, admission control, price discovery (\cref{sec:f2}).

    \item Federation handles broker death and network partition with completion rate ${\geq}\,98.7\%$ across 75 federation and resilience cells (\cref{sec:f3}).

    \item Sovereignty enforcement adds zero observable runtime overhead. Across 60 governance-grid runs (four enforcement scenarios $\times$ three pipeline types $\times$ five seeds), mean latency under sovereignty constraints matches the unconstrained baseline to within 1--3\,ms, demonstrating that the autonomic substrate honours domain-level invariants without external orchestration cost (\cref{sec:f4}).
\end{enumerate}

\subsection{Infrastructure Validation}\label{sec:infra-results}

Before the main findings, we establish four substrate-soundness results. (i)~\emph{Transport orthogonality}: at $\lambda{=}5$\,pps single-VM, per-seed latency differences between HTTP and Kafka are $<1$\,ms for round-robin and random placements; neural placement adds ${\sim}8$\,ms under Kafka relative to HTTP (Kafka consumer dispatch, $<2\%$ of total). All strategies achieve near-identical performance under homogeneous single-site conditions (${\sim}1231$\,ms median, $\geq 99\%$ completion). (ii)~\emph{Slice-aware placement} reduces median latency by 15\% under HTTP (1233 vs 1450\,ms) and increases throughput $5{\times}$ in the equalised B1eq-vs-B2 comparison. (iii)~\emph{Governance enforcement} adds 0\% latency overhead vs unconstrained slice-aware placement (within 1\,ms across HTTP and Kafka). (iv)~\emph{Worker-failure recovery}: HTTP achieves 100\% pipeline completion at $t=5$\,min worker kill; Kafka shows 1.1\% completion loss attributable to consumer-group rebalancing rather than the placement algorithm. The 4-VM baseline (40 runs) and slicing (50 runs) phases reproduce these confirmations on the production substrate at 100\% completion across all strategies and transports, with the neural-overhead vs round-robin gap shrinking to ${\approx}6$\,ms (0.5\%) and the HTTP-vs-Kafka mean-latency gap to $\leq 1$\,ms across all five placements; governance enforcement reproduces the zero-overhead result across all 50 slicing cells.

\subsection{Decentralisation Beats the Centralised Baseline}\label{sec:f1}

The efficiency comparison is:
\begin{equation}\label{eq:efficiency-gap}
    \Delta_{\text{eff}} = 1 - \frac{\eta_{\text{market}}}{\eta_{\text{oracle}}},
\end{equation}
where $\eta$ is social welfare (sum of pipeline values minus placement costs). With pipeline values held constant across strategies, $\eta$ is monotone in mean end-to-end latency, so we report mean latency directly.

The fair-process-count sharded comparator is a 4-broker centralised orchestrator with a designated coordinator (VM1, \texttt{IS\_COORDINATOR=true}) that pulls peer state via HTTP and runs \texttt{find\_placement} over the merged topology; state-owners (VMs 2--4) expose \texttt{/sharded-oracle/state}. It carries the same broker-process count as the four-broker market and produces global-optimum decisions like the single-process oracle. \Cref{tab:near-optimality} reports the pipeline-aggregated mean end-to-end latency for the four-broker market and the four-shard centralised oracle (5~seeds per cell, ${\sim}8000$--41,000 events per cell, three load levels per pipeline). The full per-cell detail (single-process oracle, four-broker market, four-shard oracle across the 9 (pipeline, load) configurations) is reported in the supplementary material as Table~S1.

\begin{table}[t]
\centering
\caption{Pipeline-aggregated mean end-to-end latency (ms) for the four-broker market and the fair-process-count four-shard centralised oracle (each entry averages 3 load levels $\times$ 5 seeds). $\Delta$ is market$-$sharded in milliseconds (negative = market faster). \emph{Cell wins} counts the per-cell market/sharded/tie outcomes (tie threshold $\pm1$\,ms; 3 load levels per pipeline). The full per-cell single-process oracle / market / sharded comparison is in Table~S1.}
\label{tab:near-optimality}
\small
\setlength{\tabcolsep}{4pt}
\begin{tabular}{@{}lrrrr@{}}
\toprule
Pipeline & Market mean & Sharded mean & $\Delta$ & Cell wins (mkt/OS/tie) \\
\midrule
cqi-chain (tree)      & 1822 & 1848 & $-26$ ($1.4\%$) & 3 / 0 / 0 \\
anomaly-sp (SP)       & 1201 & 1201 & tied            & 1 / 2 / 0 \\
ran-entangled         &  965 &  971 & $-7$ ($0.7\%$)  & 1 / 1 / 1 \\
\bottomrule
\end{tabular}
\end{table}

Across the 9~(pipeline, load) cells $\times$ 5 seeds the four-broker market trades within $\pm1.5\%$ of the four-shard centralised oracle on every cell: market dominates cqi-chain (3/3 cells, 1.4\% mean), the sharded oracle is marginally favoured on anomaly-sp at the cell level (one market win, two sharded wins, all within $\pm5$\,ms), and ran-entangled splits one market win / one sharded win / one tie. Per-cell totals over the 9 cells are 5~market wins / 3~sharded wins / 1~tie. \emph{Self-organisation pays at most a 1.5\,\% coordination cost across 9 cells, dominates linear-chain by 1.4\,\%, and trades within $\pm 5$\,ms on series-parallel and entangled DAGs.} The decentralised-vs-centralised inversion observed against the single-process oracle (a ${\sim}3\%$ market gap with 45 of 45 per-seed wins, Table~S1) therefore holds against the fair-process-count comparator on tree pipelines; on series-parallel and entangled DAGs the architectural advantage becomes pipeline-structure-conditional rather than uniform.

\begin{figure}[t]
    \centering
    \begin{subfigure}[b]{0.48\columnwidth}
        \centering
        \includegraphics[width=\linewidth]{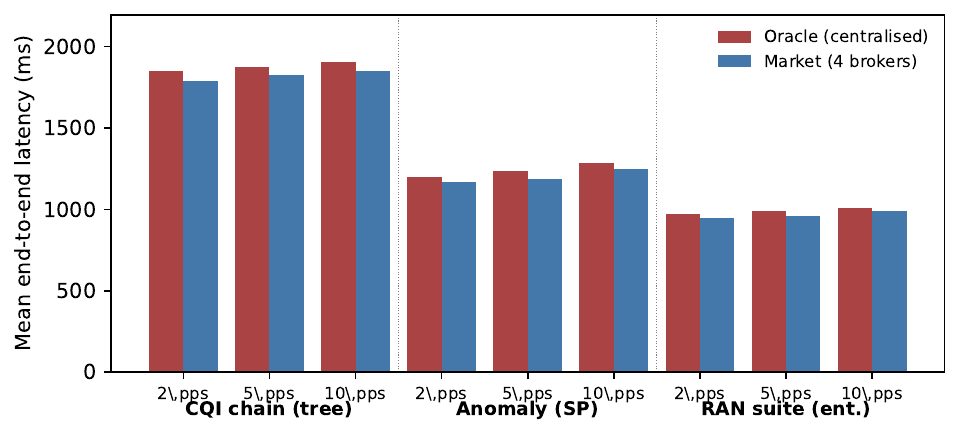}
        \caption{Near-optimality: market vs single-process oracle, all cells.}
        \label{fig:near-optimality}
    \end{subfigure}\hfill
    \begin{subfigure}[b]{0.48\columnwidth}
        \centering
        \includegraphics[width=\linewidth]{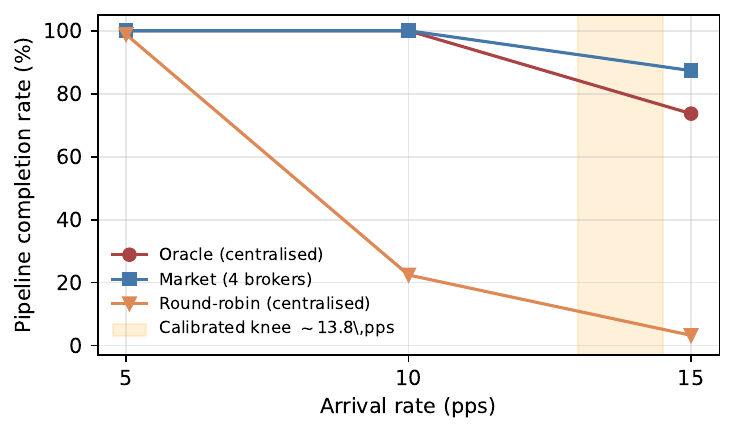}
        \caption{Round-robin CR collapse $98.8\%\to 22.4\%\to 3.3\%$.}
        \label{fig:rr-collapse}
    \end{subfigure}\\[0.6em]
    \begin{subfigure}[b]{0.48\columnwidth}
        \centering
        \includegraphics[width=\linewidth]{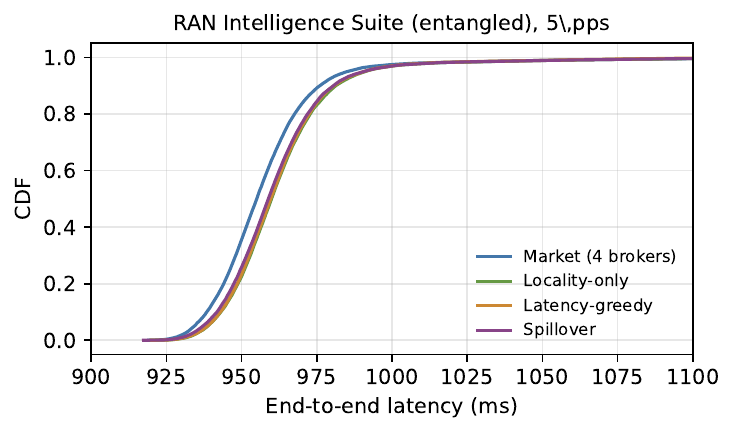}
        \caption{Heuristic strategy parity under uniform load.}
        \label{fig:heuristic-parity}
    \end{subfigure}\hfill
    \begin{subfigure}[b]{0.48\columnwidth}
        \centering
        \includegraphics[width=\linewidth]{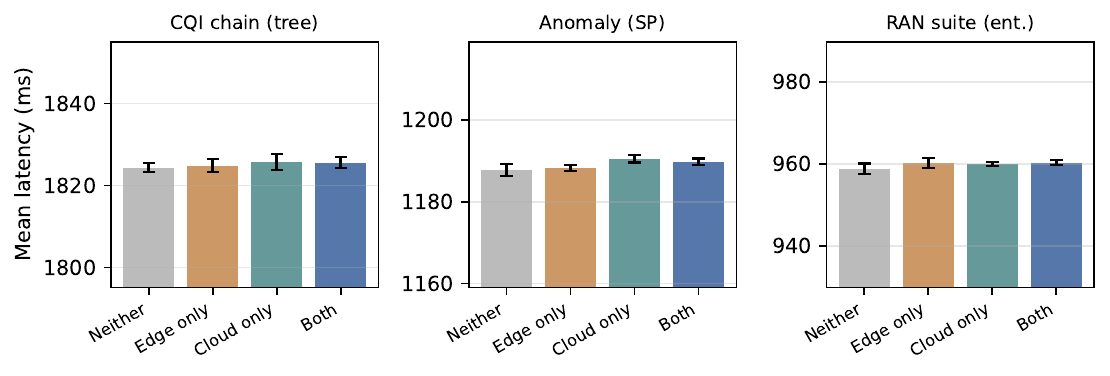}
        \caption{Sovereignty-enforcement grid: zero runtime overhead across four scenarios.}
        \label{fig:governance-grid}
    \end{subfigure}
    \caption{Consolidated experimental results: (a)~near-optimality (market vs single-process oracle, all cells); (b)~round-robin completion-rate collapse $98.8\%\to 22.4\%\to 3.3\%$ across $\lambda\in\{5,10,15\}$\,pps; (c)~heuristic strategy parity under uniform load (CDFs of mean latency for the four decentralised strategies); (d)~sovereignty-enforcement grid showing zero runtime overhead across four enforcement scenarios.}
    \label{fig:multi-panel-results}
\end{figure}

The four-broker market is uniformly faster than the single-process centralised oracle by 2--4\% in mean latency across all 9 (pipeline, load) configurations, with the market winning in every individual seed-level run (45 of 45 cells; tie threshold $|\bar{L}_{\text{market}}-\bar{L}_{\text{oracle}}|<1$\,ms, below the within-VM clock noise floor); the per-cell detail underpinning this count is in Table~S1. \emph{This inverts the classical centralised-versus-decentralised comparison at the single-broker scale}: the single-process oracle becomes a contention bottleneck for state, queue depth, and dispatch decisions even at 2\,pps, while the four federated market brokers distribute these costs. The fair-process-count comparator (\cref{tab:near-optimality}) controls for this asymmetry: at equal broker-process count the gap stays within $\pm1.5\%$ across all 9 cells. The architectural advantage becomes pipeline-structure-conditional: market dominates the linear-chain cqi-chain (3/3 cells, 1.4\%), trades within $\pm5$\,ms on series-parallel anomaly-sp (1 market / 2 sharded / 0 ties; sharded is marginally favoured at +5\,ms / +2\,ms on the low-load and medium-load cells), and ties on entangled ran-entangled at the cell level (1/1/1). Completion-rate evidence corroborates the single-process-bottleneck explanation: at 10\,pps cqi-chain, single-process oracle completion drops to 99.4\% while both market and sharded oracle remain at 100.0\%, the only main-grid configuration where any strategy lost completeness.

\emph{Statistical anchor.} A one-tailed paired sign-test on the 45-cell single-process oracle comparison gives $p = 2^{-45}\approx 2.8\times 10^{-14}$ for 45 of 45 directional wins. The Hodges--Lehmann median of the 45 per-seed paired differences (market $-$ oracle) is $-39.6$\,ms with a percentile bootstrap 95\% CI of $[-44.1,\,-35.3]$\,ms ($B=10{,}000$ resamples), confirming a robust market advantage well-bounded away from zero. The directional signal is consistent at p50, p95, and p99 across all 9 cells: the market leads the single-process oracle at every reported percentile, with one cqi-chain-high p99 cell within run-to-run noise (Suppl.\ Tab.~S2).

\emph{Convergence and pipeline structure.} The bounded staleness $B\approx 10.05$\,s (\cref{sec:consistency}) sets a price-signal-lag ceiling: up to $\lambda B\approx 139$ pipelines may arrive at one broker between exchanges; the within-round reservation ledger (\cref{sec:market-impl}) absorbs this transient locally because over-allocation onto a single worker is bounded by a single pipeline's stage count, and the next exchange resynchronises peer prices. This explains why the decentralisation inversion holds at the $\lambda\le 10$\,pps loads we test rather than the $\lambda>13.8$\,pps regime where staleness would dominate. The pipeline-structure dimension (tree, SP, entangled) does not change the picture against the single-process oracle: the gap is uniform across all three structures. The classical prediction that entangled DAGs would widen the gap (cross-tree complementarities) is not visible. We read this finding as evidence that integrator encapsulation (\cref{sec:integrator-concept}) recovers a polymatroidal quotient graph for all three structures: under encapsulation the tree, SP, and entangled DAGs collapse to the same agent-facing feasibility region, and the empirical structural axis is therefore vestigial in this regime. An alternative reading, that the within-round reservation ledger suppresses over-commitment that would otherwise translate complementarities into efficiency loss, is not separately tested by the present campaign; disambiguating the two readings requires measuring the empirical non-modularity gap $\gamma$ pre- and post-encapsulation and is left to future work.

\subsection{Strategy Differentiation Requires Stress}\label{sec:f2}

Under uniform conditions in the main allocation campaign, the market and three heuristic baselines (locality-only, latency-greedy, spillover) converge to within ${\pm}5$\,ms of each other across every (pipeline, load) cell, and locality-only is competitive with market across all 9 cells; the per-cell detail is in Table~S3, with the corresponding latency CDF in \cref{fig:heuristic-parity}. The classical centralised round-robin orchestrator is also competitive at 5\,pps. The 450-run ablation programme exposes the conditions under which the market mechanism actually differentiates: Table~S5 reports the conventional centralised round-robin orchestrator across three stress dimensions on cqi-chain pipelines.

\textbf{Saturation (\cref{fig:rr-collapse}).} The conventional round-robin orchestrator drops from CR\,=\,98.8\% at $\lambda{=}5$\,pps to 22.4\% at $\lambda{=}10$\,pps to 3.3\% at $\lambda{=}15$\,pps, with mean latency rising from 1.8\,s to 9.6\,s. The transition is sharp around the calibrated saturation knee at $\lambda{\approx}13.8$\,pps. The market mechanism, by contrast, holds CR\,=\,100\% at $\lambda \le 10$\,pps and degrades gracefully thereafter (83.3\% at $\lambda{=}15$\,pps; 25.4\% at $\lambda{=}50$\,pps where rr-global completes 0\%). The fair-process-count sharded oracle tracks the market closely across the entire sweep: 1832\,ms / 100\% at sat-5, 1853\,ms / 100\% at sat-10, and 2105\,ms / 83.8\% at sat-15 (full $n=5$ seeds for sharded; market partial $n=4$ at sat-15). At the saturation regime ($\lambda{=}15$\,pps) market and sharded oracle differ by 14\,ms in mean latency and 0.5 percentage points in CR — both autonomic-coordinated centralised and decentralised architectures degrade gracefully and identically when the testbed crosses the calibrated knee, while \texttt{rr-global} collapses. The single-process oracle's sat-15 cell is partial ($n=2$) but tracks worse on CR (74.1\%) than either four-broker mechanism. Table~S4 reports the full saturation sweep across all four strategies.

\textbf{Failure $\times$ load interaction.} Under worker kill at mid-measurement, rr-global's CR depends sharply on offered load: at 5\,pps it survives kill of 12 workers (96.7\% CR); at 8\,pps it collapses (33.5\%); at 10\,pps further (22.5\%). The market mechanism's clearing prices encode worker availability immediately (a dead worker has effectively infinite price), and its CR remains 100\% across all scenarios.

\textbf{Heterogeneity.} With edge workers $2{\times}$ slower and cloud workers $1.5{\times}$ faster, rr-global produces 2508 / 1544 / 1251\,ms mean latency on cqi-chain / anomaly-sp / ran-entangled while the market produces 1290 / 851 / 702\,ms — 48.6\% / 44.9\% / 43.9\% reductions ($\approx 46\%$ mean). Round-robin bottlenecks on the slow edge tier; the market's speed-scaled bid (\cref{sec:market-impl}) routes preferentially to fast cloud workers when they have capacity.

\textbf{Three-property decomposition.} Each ablation scenario isolates a distinct Walrasian property absent by construction from non-autonomic baselines (\cref{sec:continuum-implications}): (i)~saturation tests \emph{admission control} (price-based rate limiting bounds queue length, vs unconditional round-robin dispatch); (ii)~failure-times-load tests \emph{information completeness} (price encodes worker availability immediately, vs the round-robin health-check loop); (iii)~heterogeneity tests \emph{price discovery} (scarce workers become expensive at equilibrium, vs identical treatment). These three properties are independent of GS (\cref{prop:walrasian}) and hold for any market-clearing mechanism over a polymatroidal feasibility region, jointly explaining why the market is the only decentralised strategy that survives the ablation.

\subsection{Federation Robustness}\label{sec:f3}

The 25-cell federation phase exercises broker process kill, emulated edge--cloud network partition, and the governance, neural, and static placement variants. Across all 25 runs CR remains $\geq 98.7\%$ (broker-kill, network-partition, neural, governance: 100\%; static: 98.7\%) with mean latency within 25\,ms of baseline (1834\,ms). The 50-cell worker-failure resilience phase reinforces this at the per-stage level: across both edge sites (eMBB-kill on VM2, URLLC-kill on VM1), neural-placement (S3) worker death mid-experiment yields CR $\geq 99.9\%$ across 36{,}098 events (VM1) and 36{,}092 (VM2). All resilience cells run cqi-chain (a tree, no funnel structure), so the funnel-mode failure handling variants (wait/proceed/abort) are not exercised here; we mark this as a coverage limitation in \cref{sec:threats}.

\subsection{Sovereignty Enforcement Adds Zero Runtime Overhead}\label{sec:f4}

Across all 60 governance-grid runs (4 enforcement scenarios $\times$ 3 pipeline types $\times$ 5 seeds; per-cell detail in Table~S6, illustrated in \cref{fig:governance-grid}) the mean latency matches the corresponding non-governance baseline to within 1--3\,ms, consistent with the slicing-phase governance-overhead finding (\cref{sec:infra-results}). Sovereignty enforcement (stages tagged \texttt{local-only} cannot leave the originating domain) imposes no measurable runtime cost on this scale of testbed across all four enforcement regimes (none, edge-only, cloud-only, both). This confirms that the autonomic substrate honours domain-level data-sovereignty invariants without external orchestration cost: the polymatroidal feasibility region of \cref{eq:governance} accepts arbitrary per-domain locality predicates as coordinate-wise upper bounds, and the broker's marginal-cost clearing automatically routes around forbidden assignments.

\subsection{Threats to Validity}\label{sec:threats}

\noindent\textbf{Internal validity.} Pipeline processing times are simulated (configurable delays calibrated to representative NWDAF workloads), not real ML inference. Semantic matching is exercised as a frozen, pre-calibrated component (\cref{sec:neural-router-recap}), isolating distribution-architecture variables from matching-quality variation. Market clearing uses marginal-cost pricing, a simplification of the full ascending auction of \cref{prop:walrasian}; the gap between simplified and optimal pricing is part of the measured $\Delta_{\text{eff}}$ and works in the unexpected direction in the single-process oracle comparison (\cref{sec:f1}).

\noindent\textbf{External validity.} The testbed uses 4~VMs with 48~workers across 4~administrative domains, connected by 1~emulated WAN link. All VMs reside in the same data centre (5GTNF, Oulu); the WAN characteristics are emulated via \texttt{tc qdisc netem} as a fixed 50\,ms delay plus 5\,ms uniform jitter. Bandwidth limits, packet loss, packet reordering, path asymmetry, and cross-traffic contention are out of scope for this campaign. The control-plane price-signal exchange is small (a few KB per peer per $\delta_{\text{prop}}$ epoch), so bandwidth limits are not load-bearing within the tested regime; the orchestration result is conditional on loss / reorder / asymmetry / contention being below the levels at which broker-side TCP retransmission and the $k_{\text{miss}}=3$ failover regime begin to misclassify peers as unhealthy. Cross-site validation on real WAN infrastructure (Oulu$\leftrightarrow$Tokyo, ${\sim}150$\,ms RTT) is planned as future work. The three 8-stage pipeline types span the structural spectrum (tree, SP, entangled) relevant to the polymatroidal efficiency result.

\noindent\textbf{Coverage limitations.} The funnel-mode failure-handling variants (wait, proceed, abort) defined in the architecture (\cref{sec:failures}) are not exercised by the current resilience phase, because all resilience cells use cqi-chain pipelines (a tree, no funnel structure). Funnel-pipeline experiments are scheduled for a follow-on campaign. Heterogeneous-domain regimes that would exercise multi-level governance composition theory (the subject of forthcoming companion work; asymmetric trust, capacity, or data quality across sites) are out of scope for this homogeneous-load campaign and are scheduled for follow-on work. The "edge-HPC/cloud continuum" scope advertised by the SI is exercised at the orchestration layer (four-domain federation, sovereignty enforcement, cross-site WAN emulation) but not at the infrastructural layer (single data centre, no HPC tier); see the Scope paragraph in \cref{sec:intro}.

\noindent\textbf{Construct validity.} The efficiency measure $\eta$ combines pipeline completion, latency, and cost. The sovereignty-enforcement scenarios (edge-only, cloud-only) test enforcement at the site level, not the individual-domain level; the 2$\times$2 design controls for site-level asymmetry. Docker containers share VM CPU and memory; CPU pinning mitigates resource contention between brokers. The five engineering refinements of \cref{sec:market-impl} are jointly sufficient for the empirical envelope tested; per-component leave-one-out attribution is deferred to follow-on work.
            %% §5 Evaluation (testbed + four findings)

\section{Discussion}\label{sec:Discussion}

\subsection{Security and Governance}\label{sec:security}\label{sec:limitations}

\emph{Limitations.} The 4-VM evaluation uses four logical domains (48 workers, four brokers); production deployments scale to hundreds of workers per domain. \cref{prop:walrasian} is scale-independent, but $\Delta_{\text{eff}}$ may differ at larger scale. Pipeline processing times are configurable delays rather than real ML inference, by design to isolate placement / market mechanisms as independent variables. The marginal-cost pricing of \cref{alg:market} approximates the polyhedral clinching auction of \cref{prop:walrasian}; the gap is part of the measured efficiency gap. The full threats-to-validity analysis is in \cref{sec:threats}.

The Neural Pub/Sub broker operates in embedding space and exchanges only subscription summaries (centroid embeddings + cluster radii) and price signals, neither revealing individual subscription text or worker-level cost information; embedding-inversion risks~\cite{morris2023embedinv} qualify this — text-embedding inversion attacks can recover fragments of source text from raw embeddings, so privacy ``reduces but does not eliminate raw-content leakage''. In adversarial settings, brokers may inflate prices or deflate capacity; the current architecture assumes cooperative brokers under a single operator. Commitment-based mechanisms grounded in polyhedral clinching auctions~\cite{ausubel2004ascending,goel2015polyhedral} (already invoked in Section~S2 of the electronic supplement) and post-hoc evidence-bundle auditability are well-known directions for the multi-vendor adversarial regime; their integration is outside this paper's scope.

Sovereignty enforcement adds zero observable overhead (\cref{sec:f4}); sovereignty constraints are motivated by data-protection regulations (e.g., GDPR data residency) but the substrate accepts arbitrary per-domain locality predicates as coordinate-wise bounds on the polymatroidal feasibility region (\cref{sec:formal}). Forthcoming companion work on multi-level governance composition predicts that partial enforcement across credibility levels can be structurally counterproductive under asymmetric trust, capacity, or data quality; testing this in a heterogeneous-domain campaign is future work and lies outside the homogeneous-load regime studied here.

\subsection{Deployment Gradient and Continuum Self-Management}\label{sec:deployment-gradient}\label{sec:continuum-implications}

The polymatroidal allocation framework is currency-agnostic~\cite{trilogy_p1}, enabling a three-level deployment gradient: \emph{(i) internal priority scheduling} (single domain, synthetic priority tokens, minimal governance — no real economy needed), \emph{(ii) federated priority} (cross-domain, shared priority-token price signals via \cref{sec:market-allocation}), and \emph{(iii) full service economy} (real currency, commitment-based credibility stack outside this paper's scope). Disaggregated continuum deployments (e.g., O-RAN) create precisely the multi-domain scenario the substrate addresses; the price-signal layer adds an economic-incentive layer (price-signal correctness, under-/over-reporting accounting) complementing cryptographic trust controls (mTLS / OAuth / certificate-based identity, e.g., O-RAN WG11~\cite{oranwg11sec}). The market preserves operator/vendor confidentiality (each broker exposes only aggregate price signals; internal cost structures and capacity utilisation remain private), while a centralised orchestrator would require disclosing proprietary operational details, foreclosing adoption. The single-process oracle's high-load sensitivity (\cref{sec:f1}) echoes the ``reallocation storm'' phenomenon~\cite{loven2021weathering,loven2022stormy}; the price-signal feedback loop dampens it (overloaded domains become expensive and divert traffic). The three Walrasian properties of \cref{sec:f2} (information completeness, admission control, price discovery) are independent of GS and delineate when round-robin suffices (uniform load, single visibility scope) versus when a market is required (multi-domain continuum, asymmetric capacity / failure / trust, vendor confidentiality). Production single-cluster schedulers (Kueue~\cite{kueue2024}, Volcano~\cite{volcano2024}) recover admission control via priority queues with quotas, but price discovery and federation-level information completeness presuppose multi-control-plane visibility that single-cluster schedulers do not provide. Multi-cluster federation operators (KubeFed, Karmada, Liqo) and cross-cluster cohort-borrowing extensions (Kueue MultiKueue) close the multi-control-plane gap at the orchestration layer, but on a different axis: they rely on a federated control-plane operator that has visibility into every cluster's state, whereas Neural Pub/Sub brokers exchange only aggregate price signals and subscription summaries with bounded staleness, preserving operator confidentiality. The substrate also commits at per-pipeline timescales (one clearing decision per pipeline arrival) rather than the per-pod / per-job timescales of cluster-federation operators. Scaling to hundreds of workers per domain via hierarchical broker federation has $O(m\cdot\bar{p})$ federation cost (\cref{sec:scaling}); validation via EISim~\cite{eisim2023} is future work.

\emph{HPC-tier integration.} The Special Issue spans the edge--\emph{HPC}/cloud continuum~\cite{beckman2020harnessing,parashar2025everywhere}; the present campaign exercises the edge--cloud tiers but not a tier-1 supercomputing site. Architecturally, an HPC domain enters the substrate as a batch-queue endpoint rather than a worker pool, and the federation contract degrades gracefully to it: integrator encapsulation (\cref{sec:integrator-concept}) contracts the HPC sub-DAG to a single composite node whose marginal-cost bid is the queue's expected wait, so a batch-tier domain participates in market clearing as one coarse-grained composite worker without exposing its internal scheduler. The within-round reservation ledger (\cref{sec:market-impl}), however, is calibrated to the ${\sim}10$\,s bounded-staleness regime of \cref{sec:f1}: bulk-synchronous HPC queueing introduces wait-time variance at the $10^2$--$10^3$\,s timescale, at which a within-round ledger is too coarse to track admission and a coarser commitment epoch (one clearing per batch window rather than per pipeline arrival) is required. Quantifying that regime --- a Pegasus-style cross-data-centre workflow with a tier-3 site --- needs an HPC testbed arm and is deferred to the follow-on systems campaign; the autonomic mechanism is HPC-tier-compatible by construction but unproven in that regime.

\subsection{Assurance and Verification}\label{sec:assurance}

The autonomic loop's correctness rests on a stated operating envelope rather than a closed-form assurance proof: bounded staleness $B = \delta_{\text{prop}} + \text{WAN}_{\max}\approx 10.05$\,s, arrival rate $\lambda \le 13.8$\,pps for the round-robin saturation knee with $\lambda B \le 139$ pipelines absorbed by the within-round reservation ledger, and federation peer-liveness misclassification bounded by the $k_{\text{miss}}=3$ failover regime under the $\delta_{\text{prop}}=10$\,s exchange period (\cref{sec:consistency,sec:market-impl}). Within this envelope, the substrate's safety invariants are: (i) no pipeline stage is dispatched to a worker with $\rho_i \ge 0.99$ (admission-control guarantee from the $\rho_i$ cap in \cref{eq:mm1-cost}), (ii) no within-round over-commitment exceeds a single pipeline's stage count (reservation-ledger invariant), and (iii) sovereignty constraints are coordinate-wise upper bounds on the feasibility region and cannot be violated by any market-clearing decision (\cref{sec:formal}). Liveness rests on the eventual-consistency reading of \cref{prop:walrasian}: each broker's local clearing converges to the equilibrium price under bounded staleness as the next federation exchange resynchronises peer prices. Statistical model-checking of these invariants under perturbation~\cite{calinescu2012qosaware,tamura2013runtimemodels,weyns2020software} would strengthen the assurance argument and is a natural next step; the present paper substantiates the envelope empirically through the perturbation campaign of \cref{sec:Evaluation} (failure injection, partition, saturation sweeps) rather than through formal verification.

\subsection{Open Questions}\label{sec:open}

Three directions remain open. \emph{Broker incentive alignment}: federated brokers face strategic incentives (price misreporting, capacity withholding, cross-domain favouritism) under adversarial multi-vendor deployments; broker accountability (commitment-based pricing, cross-broker auditing, coalition deterrence) integrated with the federation protocol is future work. \emph{Dynamic credibility and multi-agent negotiation}: governance enforcement adapting to evolving trust levels via repeated-game analysis (subject of forthcoming companion work), and broker price signals as a coordination layer beneath MAS negotiation protocols. \emph{Semantic-communication transport}: semantic codecs~\cite{xie2021deep,qin2022semantic} beneath Neural Pub/Sub's application-layer routing (\cref{sec:semcomm-positioning}) could reduce backhaul traffic while preserving application-level matching; quantifying the bandwidth saving is future work.

            %% §6 Discussion (autonomic implications, limitations)
\section{Conclusion}\label{sec:Conclusion}

We presented \emph{Neural Pub/Sub}, an autonomic federated-broker substrate for the edge--cloud computing continuum, in which self-organisation arises from market-based price signals rather than centralised control. The substrate closes a MAPE-K control loop at each broker over per-worker monitoring, marginal-cost clearing-price analysis, polymatroidal placement planning, federated cross-domain dispatch, and bounded-staleness shared knowledge (\cref{sec:mapek}). The design is anchored in a Walrasian convergence proposition imported from companion work (\cref{prop:walrasian}, Section~S2 of the electronic supplement): under gross-substitutes valuations on tree and series-parallel service-dependency DAGs, decentralised price-based allocation matches the welfare of a centralised oracle.

A 1005-run, seven-phase campaign on a 4-domain, 48-worker testbed, supplemented by a fair-process-count sharded-oracle comparator, supports four findings. The four-broker market beats a single-process oracle by 2--4\% in mean latency across all 9 (pipeline, load) cells with 45 of 45 per-seed wins (sign-test $p\approx 2.8\times 10^{-14}$); against the fair-process-count four-shard centralised oracle the gap is within $\pm1.5\%$ across all 9 cells, with the market dominating linear-chain cqi-chain (3/3) and ties or marginal sharded wins on series-parallel anomaly-sp and entangled ran-entangled (\cref{sec:f1}). Self-management differentiates only under stress: round-robin completion rate collapses $98.8\%\to 22.4\%\to 3.3\%$ across $\lambda\in\{5,10,15\}$\,pps while the market preserves CR, and the advantage decomposes into three Walrasian properties (information completeness, admission control, price discovery) absent from non-autonomic baselines (\cref{sec:f2}). The federation withstands broker process kill, emulated cross-site partition, and per-stage worker death across both edge sites with CR\,$\geq\,98.7\%$ across 75 cells (\cref{sec:f3}). Sovereignty enforcement adds zero measurable runtime overhead across 60 governance-grid runs (four enforcement scenarios $\times$ three pipeline types $\times$ five seeds), confirming that the autonomic substrate honours domain-level data-sovereignty invariants without external orchestration cost (\cref{sec:f4}).

The contributions are an autonomic federated-market substrate for the edge--cloud continuum, the empirical decentralised-vs-centralised inversion at equal process count, and a structural decomposition of when autonomic strategies differ. Future work targets heterogeneous-domain experiments designed to test multi-level governance composition theory (the subject of forthcoming companion work) under asymmetric trust, capacity, and data quality, scaling validation via EISim~\cite{eisim2023} to production-scale deployments, broker-accountability mechanisms (commitment-based pricing, cross-broker auditing) for adversarial multi-vendor settings, and dynamic credibility adaptation. The substrate's currency-agnostic design supports a deployment gradient from internal priority scheduling within a single domain to full service economies spanning multiple administrative domains and jurisdictions, with the same MAPE-K guarantees applying at each stage.
            %% §7 Conclusion

\section*{ACKNOWLEDGMENTS}\label{sec:acknowledges}
This work was supported by the Research Council of Finland through the 6G Flagship programme (grant 318927), the Strategic Research Council affiliated with the Academy of Finland through the CO2CREATION project (grant 372355), by Business Finland through the Neural pub/sub research project (diary number 8754/31/2022), and by the European Regional Development Fund (ERDF; project numbers A81568, A91867).

\section*{Conflict of Interest}
The authors declare no competing financial or personal interests that could have appeared to influence the work reported in this paper.

\section*{Code and Data Availability}\label{sec:availability}
The broker codebase, run-driver scripts, per-cell raw CSV data, and analysis notebooks underlying the paper's results are released at \url{https://github.com/lloven/neural-pubsub-experiments} under the Apache-2.0 license and archived at Zenodo (\url{https://doi.org/10.5281/zenodo.20392551}); see \texttt{REPRODUCING.md} (seven-phase campaign protocol, smoke tests, figure scripts) and \texttt{CITATION.cff} in the repository.

%% ===== Electronic supplement folded in (single-file arXiv build) =====
\clearpage
\setcounter{section}{0}\setcounter{subsection}{0}\setcounter{table}{0}\setcounter{figure}{0}\setcounter{equation}{0}
\renewcommand{\thesection}{S\arabic{section}}
\renewcommand{\thesubsection}{\thesection.\arabic{subsection}}
\renewcommand{\thetable}{S\arabic{table}}
\renewcommand{\thefigure}{S\arabic{figure}}
\renewcommand{\theequation}{S\arabic{equation}}
\section{Overview}\label{sec:supp-overview}
%% This file is the body of supplement.tex (a SEPARATE document from main.tex).
%% It is included from supplement.tex as its main content. Do not also \input it
%% from main.tex.

This electronic supplement collects the per-cell detail tables supporting the four headline findings of the main paper, whose body presents only the aggregated results; the detail here supports reviewers seeking finer-grained traceability. The supplement's own tables carry an \texttt{S}\,prefix (Tab.\,S1, etc.) to distinguish them from the main paper's.

\subsection{Per-Cell Detail: Single-Process Oracle, Market, and Sharded Oracle}\label{app:f1-detail}

\Cref{tab:near-optimality-detail} reports the full 9-cell comparison underpinning the main paper's aggregated near-optimality table: mean end-to-end latency for the single-process centralised oracle (Oracle$_1$), the four-broker market, and the four-shard centralised oracle (Oracle$_4$, designated coordinator at VM1, peer state pulled via HTTP, global-optimum decisions over the merged topology) across the 9~(pipeline, load) configurations of the main market campaign (5~seeds each, ${\sim}8000$--41,000 events per cell). The market beats the single-process oracle in every cell (45/45 per-seed wins). Against the fair-process-count sharded oracle the gap is within $\pm1.5\%$ across all 9 cells.

\begin{table}[t]
\centering
\caption{Mean end-to-end latency (ms) for single-process oracle (Oracle$_1$), four-broker market, and four-shard centralised oracle (Oracle$_4$) across pipeline structure and load. Gap columns are market$-$named-oracle in milliseconds (negative = market faster); SE is the standard error of the M$-$O$_{1}$ paired difference across the 5 seeds in the cell.}
\label{tab:near-optimality-detail}
\small
\setlength{\tabcolsep}{4pt}
\begin{tabular}{@{}llrrrrrr@{}}
\toprule
Pipeline & Load & Oracle$_{1}$ & Market & Oracle$_{4}$ & M$-$O$_{1}$ & SE & M$-$O$_{4}$ \\
\midrule
cqi-chain (tree) & low (2\,pps) & 1852 & 1792 & 1806 & $-60$ & 0.8 & $-14$ \\
cqi-chain & medium (5\,pps) & 1874 & 1824 & 1838 & $-50$ & 1.7 & $-14$ \\
cqi-chain & high (10\,pps) & 1908 & 1852 & 1901 & $-56$ & 11.8 & $-49$ \\
anomaly-sp (SP) & low & 1199 & 1170 & 1165 & $-29$ & 0.4 & $+5$ \\
anomaly-sp & medium & 1237 & 1186 & 1184 & $-51$ & 0.6 & $+2$ \\
anomaly-sp & high & 1285 & 1246 & 1253 & $-39$ & 1.9 & $-7$ \\
ran-entangled & low & 973 & 948 & 946 & $-25$ & 0.6 & $+2$ \\
ran-entangled & medium & 992 & 958 & 959 & $-34$ & 0.8 & $-1$ \\
ran-entangled & high & 1008 & 988 & 1010 & $-20$ & 2.6 & $-22$ \\
\bottomrule
\end{tabular}
\end{table}

\Cref{tab:tail-latency-detail} reports tail-latency percentiles (p50, p95, p99) for the four-broker market and the single-process oracle on the same 9~(pipeline, load) cells (events pooled across 5~seeds; success-only). Market dominates at every percentile on cqi-chain (low / medium) and on every anomaly-sp and ran-entangled cell; on cqi-chain high the p99 difference is within $+23$\,ms and within run-to-run noise (the 99th-percentile estimate fluctuates by ${\pm}30$\,ms across seeds at ${\sim}40{,}000$ events per cell). The directional signal of \cref{tab:near-optimality-detail} is consistent at p50 / p95 / p99 with the mean-latency comparison.

\begin{table}[t]
\centering
\caption{Tail-latency percentiles (ms) for the single-process oracle (Oracle$_1$) and four-broker market across the 9~(pipeline, load) cells. Events are pooled across 5~seeds (success-only).}
\label{tab:tail-latency-detail}
\small
\setlength{\tabcolsep}{3pt}
\begin{tabular}{@{}llrrrrrr@{}}
\toprule
& & \multicolumn{3}{c}{Oracle$_{1}$} & \multicolumn{3}{c}{Market} \\
\cmidrule(lr){3-5}\cmidrule(lr){6-8}
Pipeline & Load & p50 & p95 & p99 & p50 & p95 & p99 \\
\midrule
cqi-chain & low    & 1841 & 1933 & 1989 & 1787 & 1829 & 1878 \\
cqi-chain & medium & 1856 & 2012 & 2102 & 1814 & 1904 & 2004 \\
cqi-chain & high   & 1859 & 2107 & 2211 & 1816 & 2070 & 2234 \\
anomaly-sp & low    & 1191 & 1256 & 1320 & 1168 & 1191 & 1235 \\
anomaly-sp & medium & 1224 & 1339 & 1433 & 1181 & 1228 & 1306 \\
anomaly-sp & high   & 1257 & 1487 & 1643 & 1223 & 1411 & 1621 \\
ran-entangled & low    & 970 & 1003 & 1055 & 947 & 962 & 980 \\
ran-entangled & medium & 984 & 1060 & 1149 & 955 & 985 & 1052 \\
ran-entangled & high   & 990 & 1137 & 1276 & 974 & 1081 & 1262 \\
\bottomrule
\end{tabular}
\end{table}

\subsection{Per-Cell Detail: Heuristic Strategies and Round-Robin Stress}\label{app:f2-detail}

Under uniform conditions in the main allocation campaign, the market and three heuristic baselines (locality-only, latency-greedy, spillover) converge to within ${\pm}5$\,ms of each other across every cell (\cref{tab:heuristic-comparison-detail}); locality-only is competitive with market across all 9 cells. The 450-run ablation programme exposes the conditions under which the market mechanism actually differentiates. \Cref{tab:rr-results-detail} reports the conventional centralised round-robin orchestrator across three stress dimensions on cqi-chain pipelines.

\begin{table}[t]
\centering
\caption{Mean latency (ms) across decentralised strategies under uniform conditions; all four converge.}
\label{tab:heuristic-comparison-detail}
\scriptsize
\begin{tabular}{@{}llrrrr@{}}
\toprule
Pipeline & Load & Market & Locality & Lat-greedy & Spillover \\
\midrule
cqi-chain & low & 1792 & 1791 & 1797 & 1794 \\
cqi-chain & med & 1824 & 1819 & 1828 & 1825 \\
cqi-chain & high & 1852 & 1845 & 1849 & 1846 \\
anomaly-sp & low & 1170 & 1170 & 1174 & 1173 \\
anomaly-sp & med & 1186 & 1189 & 1192 & 1190 \\
anomaly-sp & high & 1246 & 1240 & 1248 & 1240 \\
ran-entangled & low & 948 & 950 & 949 & 949 \\
ran-entangled & med & 958 & 963 & 962 & 961 \\
ran-entangled & high & 988 & 995 & 993 & 992 \\
\bottomrule
\end{tabular}
\end{table}

\begin{table}[t]
\centering
\caption{Saturation sweep across all four placement strategies on cqi-chain (5~seeds per cell except where indicated). The four-broker market and the four-shard centralised oracle remain within $\pm15$\,ms across every load and report identical completion rates within 0.5~percentage points; \texttt{rr-global} collapses dramatically at $\lambda \ge 10$\,pps. The single-process oracle's sat-15 cell is partial (n=2 from an earlier campaign); market-quad sat-15 is n=4.}
\label{tab:saturation-detail}
\small
\begin{tabular}{@{}lrrrr@{}}
\toprule
\multirow{2}{*}{Strategy} & \multicolumn{2}{c}{sat-5 (5\,pps)} & \multicolumn{2}{c}{sat-10 (10\,pps)} \\
 & CR\% & mean ms & CR\% & mean ms \\
\midrule
oracle (single-process) & 100.0 & 1872 & 100.0 & 1888 \\
market (4-broker) & 100.0 & 1845 & 100.0 & 1845 \\
oracle (4-shard) & 100.0 & 1832 & 100.0 & 1853 \\
rr-global & 95.5 & 1999 & 19.9 & 4803 \\
\midrule
\multicolumn{5}{c}{sat-15 (15\,pps): saturation regime} \\
\midrule
oracle (single-process) & \multicolumn{2}{c}{74.1\% / 2331\,ms (n=2)} & \multicolumn{2}{c}{} \\
market (4-broker) & \multicolumn{2}{c}{83.3\% / 2091\,ms (n=4)} & \multicolumn{2}{c}{} \\
oracle (4-shard) & \multicolumn{2}{c}{83.8\% / 2105\,ms} & \multicolumn{2}{c}{} \\
rr-global & \multicolumn{2}{c}{3.3\% / 10981\,ms} & \multicolumn{2}{c}{} \\
\bottomrule
\end{tabular}
\end{table}

\begin{table}[t]
\centering
\caption{Conventional round-robin orchestrator across three stress dimensions, on cqi-chain pipelines. The market preserves CR\,=\,100\% at every load tested up to $\lambda{=}10$\,pps and degrades gracefully thereafter (CR\,=\,25.4\% at $\lambda{=}50$\,pps where rr-global is 0\%).}
\label{tab:rr-results-detail}
\small
\begin{tabular}{@{}lllrr@{}}
\toprule
Stress dimension & Scenario & Description & rr-global CR\% & rr-global mean (ms) \\
\midrule
\multirow{3}{*}{Saturation (no failure)} & sat-5 & $\lambda{=}5$\,pps, no failure & 98.8 & 1843 \\
 & sat-10 & $\lambda{=}10$\,pps, no failure & 22.4 & 3395 \\
 & sat-15 & $\lambda{=}15$\,pps, no failure & 3.3 & 9598 \\
\midrule
\multirow{3}{*}{Failure $\times$ load} & failure-5-12 & $\lambda{=}5$\,pps, kill 12 workers & 96.7 & 1940 \\
 & failure-8-12 & $\lambda{=}8$\,pps, kill 12 workers & 33.5 & 3262 \\
 & failure-10-12 & $\lambda{=}10$\,pps, kill 12 workers & 22.5 & 3708 \\
\midrule
Heterogeneity & het & 2$\times$ slow edge / 1.5$\times$ fast cloud workers & 94.3 & 2508 \\
\bottomrule
\end{tabular}
\end{table}

\subsection{Per-Cell Sovereignty-Enforcement Grid}\label{app:f4-detail}

\Cref{tab:governance-results-detail} reports mean latency for the four sovereignty-enforcement scenarios (none, edge-only, cloud-only, both) across three pipeline types at medium load (5~seeds each, $5{\times}3{\times}4 = 60$ runs). All four scenarios are within $\pm3$\,ms of each other across every pipeline, demonstrating zero measurable runtime overhead from sovereignty enforcement (Finding~4 of the main paper).

\begin{table}[t]
\centering
\caption{Sovereignty-enforcement grid: mean latency (ms) across four enforcement scenarios. All scenarios are within $\pm3$\,ms of each other across every pipeline.}
\label{tab:governance-results-detail}
\small
\begin{tabular}{@{}lrrrr@{}}
\toprule
Pipeline & none & edge-only & cloud-only & both \\
\midrule
cqi-chain & 1824 & 1825 & 1826 & 1826 \\
anomaly-sp & 1188 & 1188 & 1191 & 1190 \\
ran-entangled & 959 & 960 & 960 & 960 \\
\bottomrule
\end{tabular}
\end{table}

% S1.5 "Threats to Validity" removed 2026-05-26 (AE review): it was a hollow
% redirect; the full internal/external/coverage/construct analysis is in the
% main paper's threats section (sec:threats).

\section{Self-Contained Statement of Proposition 3.1}\label{app:propositions}

This section of the electronic supplement restates the Walrasian convergence proposition on which the main paper's market mechanism is grounded, with conditions stated formally and a pointer to where the proof lives in companion work~\cite{trilogy_p1}. The proposition is imported as a black box: its proof is not reproduced here, but the conditions under which it holds are made explicit so that a reviewer can verify whether the conditions are met by the deployment of the main paper's distribution architecture (\S4). Cross-references of the form ``\S4'' or ``Eq.~(3)'' point into the main paper; this supplement compiles standalone.

\textbf{Setting.} Let $G_{\mathrm{res}}=(\mathcal{R},E)$ be a service-dependency DAG with capacities $\{C_v\}_{v\in\mathcal{R}}$ and leaf services $L(G)\subseteq\mathcal{R}$, defining the service-feasibility region $\mathcal{X}_{\mathrm{res}}$ of the main paper (Eq.~(3)). Let $\mathcal{X}_t=\mathcal{X}_{\mathrm{res}}\cap\mathcal{X}_{\mathrm{gov}}$ be the governance-constrained feasibility region (main paper, Eq.~(8)).

\textbf{Conditions.}
\begin{enumerate}
    \item \emph{Laminar structure.} The constraint family $\{L_v : v\in\mathcal{R}\setminus L(G)\}$ is laminar on $L(G)$: for any two members $L_v, L_{v'}$, either $L_v\subseteq L_{v'}$, $L_{v'}\subseteq L_v$, or $L_v\cap L_{v'}=\emptyset$. This holds in particular when $G_{\mathrm{res}}$ is a rooted tree or a two-terminal series-parallel network (see the main paper's \S3.1 for the parse-tree structural argument).
    \item \emph{Polymatroidal feasibility.} Under condition (1), $\mathcal{X}_{\mathrm{res}}$ is a polymatroid with rank function $f$ defined by Eq.~(4) of the main paper~\cite{edmonds1970submodular,fujishige2005submodular}. Coordinate-wise governance bounds preserve polymatroidal structure; therefore $\mathcal{X}_t$ is also polymatroidal.
    \item \emph{Gross-substitutes valuations.} Each agent's valuation over leaf-allocation slices satisfies the gross-substitutes (GS) condition of~\cite{kelso1982job}. This condition is non-trivial: pipeline-bundle valuations are perfect-complements (Leontief) on the raw stage set, and the GS-compatible representation arises only after \emph{integrator encapsulation}~\cite{trilogy_p1}, where each domain's broker bundles its multi-resource sub-DAG into a single composite slice with unit demand at the agent-facing level (main paper, \S3.1.3 and \S4.2.4).
\end{enumerate}

\textbf{Buyer-side GS sketch.} The agent's raw valuation over the eight stage types is Leontief: $v_{\text{raw}}(x)=v^{\star}\cdot\mathbf{1}[x\ge\mathbf{1}_V]$ on $\{0,1\}^V$, which is not GS. After integrator encapsulation each domain's broker contracts its sub-DAG into a composite item; the buyer-facing item set becomes the set of traversed domains $\mathcal{D}$ and the valuation $v(x)=v^{\star}\cdot\mathbf{1}[x\ge\mathbf{1}_D]$ is unit-demand-per-domain on the integrator quotient. Unit-demand satisfies GS~\cite{kelso1982job}, so the buyer-facing layer is GS by construction and the non-modularity gap $\gamma$ of the main paper's \S5.1 is $\gamma=0$ post-quotient. ``Residual encapsulation losslessness'' (the integrator's max-flow, main paper Eq.~(10), is achieved without an internal sub-DAG bottleneck, so the composite item's unit demand is realisable) is the load-bearing assumption deferred to~\cite{trilogy_p1}.

\textbf{Statement and where the proof lives.} Under conditions (1)--(3): (a)~a Walrasian equilibrium $(\bm{p}^*,\bm{x}^*)$ exists; (b)~$\bm{x}^*$ maximises social welfare $\sum_i v_i(x_i^*)$; (c)~$(\bm{p}^*,\bm{x}^*)$ is computable in polynomial time by an ascending polyhedral-clinching auction~\cite{ausubel2004ascending,goel2015polyhedral} on the polymatroidal region $\mathcal{X}_t$. Conditions (1)--(2) are classical~\cite{edmonds1970submodular,fujishige2005submodular}; (a)--(b) follow from Kelso--Crawford~\cite{kelso1982job} and Gul--Stacchetti~\cite{gul1999grosssubstitutes} given the polymatroidal structure; the AI-pipeline construction and the integrator-encapsulation reduction are in~\cite{trilogy_p1}; (c) follows from~\cite{ausubel2004ascending,goel2015polyhedral}.

\textbf{Implementation gap.} Algorithm~1 of the main paper implements marginal-cost clearing (main paper, Eqs.~(14) and~(11)), not the full polyhedral-clinching auction of (c) (cf.\ the main paper's \S4.3.2 five-component refinement); the empirical efficiency gap of the main paper's \S5.5 therefore characterises the distance between marginal-cost clearing and the welfare-maximising equilibrium of Proposition~3.1, not the equilibrium itself.

\textbf{Deployment laminarity check.} The parse-tree argument of the main paper (\S3.1.2) certifies that all three deployment pipelines satisfy condition~(1): cqi-chain is a rooted tree, anomaly-sp is two-terminal series-parallel (parallel detector branch composed with a fusion stage at the near-RT RIC), and ran-entangled becomes laminar after integrator encapsulation~\cite{trilogy_p1} contracts each domain's sub-DAG to a composite node (main paper, Eq.~(5)), yielding a 4-node tree quotient; conditions (2)--(3) then follow.

\bibliographystyle{ACM-Reference-Format}
\bibliography{references}

%% Electronic supplement (per-cell detail tables) lives in a separate compile
%% unit at supplement.tex (rendered as supplement.pdf). Cross-references to it
%% (e.g.\ ``Suppl.\ Tab.~S2'') are hard-coded; the documents compile standalone.

\end{document}